\documentclass[10pt]{article}
\usepackage{fullpage}
\usepackage[T1]{fontenc}
\usepackage{lmodern}
\usepackage{pdfrender}
\usepackage[dvipsnames]{xcolor}
\usepackage{amsmath}
\usepackage{amssymb}
\usepackage{mathrsfs}
\usepackage{mathbbol}
\usepackage{graphicx}
\usepackage[colorlinks=true, citecolor=black, linkcolor=black, urlcolor=blue]{hyperref}
\usepackage{accents}
\usepackage{cite}
\usepackage{authblk}

\usepackage{slashed}
\usepackage{simplewick}
\usepackage{feynmf}

\newcommand{\B}{\boldsymbol}
\newcommand{\trace}{\text{tr}}

\newcommand{\xdownarrow}[1]{{\left\downarrow\vbox to #1{}\right.\kern-\nulldelimiterspace}}

\newcommand{\Curl}{\text{Curl}\,}

\numberwithin{equation}{section}

\title{Micropolar continua as projective space of Skyrmions}
\author{Yongjo Lee\footnote{Email: yongjo.lee.16@ucl.ac.uk}}
\affil{\small{\textsl{Department of Mathematics, University College London},\\\textsl{Gower Street, London, WC1E 6BT, UK}}}
\date{}

\begin{document}

\maketitle
\begin{abstract}
Micropolar continua are shown to be the generalisation of nematic liquid crystals through perspectives of order parameters, topological and geometrical considerations. Micropolar continua and nematic liquid crystals are recognised as  antipodals of $S^3$ and $S^2$ in projective geometry. We show that position-dependent rotational axial fields in kinematic micropolar continua can be considered as solutions of anisotropic Higgs fields, characterised by integers $N$. We emphasise that the identical integers $N$ are topological invariants through homotopy classifications based on defects of order parameters and a finite energy requirement. Magnetic monopoles and Skyrmions are investigated based on the theories of defects of continua in Riemann-Cartan manifolds.
\end{abstract}

\mbox{}

\section{Introduction}
\subsection{Background and motivation}
If we can find a solution space for a given system, there might be a number of solutions that can be transformed continuously around the most stable solution. If these comparable and equivalent solutions form a distinct set under an internal symmetry, we might associate the set of solutions a group structure within the allowed finite energy of the system. This observation shares many similarities in describing defects of a body which can undergo smooth deformations but restores its original shape when the deformational factors are removed.

Defects of a deformable body can be classified by a set of equivalent classes when we consider a compatibility condition that is derivable from a simple integrable equation to obtain the solution space. In turn, integers might be assigned to those classifications of defects by group theoretical approaches.

The simplest example would be an assignment of an identity element $\{0\}$ to contain all configurations of classical elastic deformations. In this particular set of configuration, small fluctuations are allowed to retain the elasticity around the stable solution of the system. Other than the classical elastic regime, we can assign a class $\{1\}$ to emphasise features that differ from those of the classic elastic deformations. This kind of assignment prohibits  solution configurations of the class $\{1\}$ to continuously transform into that of the class $\{0\}$ without violating the finite energy requirement.

Instead of assigning integers by hand, we would like to see under what systematic assignment of an infinite range of integers can allow us to investigate the group classification of the large class of solution space into a discrete set of configurations. We will take two different approaches to understand the integer-valued assignment. One is based on the theory of defects in a given order parameter space and another approach is originated from the boundary conditions of field configurations. These integer-valued assignments will yield the topologically invariant quantities through various physical models.

In describing the defects in differentiable manifolds, the continuum theory contains deformational measures related to curvature and torsion. These measures are caused by broken symmetries of rotations and translations. In formulating micro or macroscopic rotations, a number of models are investigated by using a simple ansatz such as the global uniaxial field of the rotations or small rotational angles, often in one dimensional static case due to the complicated nonlinear nature of the problem. In realistic inhomogeneous settings, one eventually includes axial fields and angular variables of the rotation, depend on space and time in the given manifold, such as the Riemann-Cartan manifold, especially if one is interested in torsion and curvature at the same time. 

We would like to investigate the consequences when we consider arbitrary position-dependent axial fields of $SO(3)$, and its implications to physical systems that contain $SO(3)$ as its symmetry (sub)group in relation with the assignment of the integers when we classify the solution space.

This paper is organised as follows. In Section 1 we briefly introduce the microscopic theory of continuum physics followed by constructions of torsion and curvature tensors in Riemann-Cartan manifolds among other measures of defects. Topological and geometrical considerations of nematic liquid crystals and a definition of an order parameter space in the context of homotopy groups are given in Section 2. In Section 3, we consider links between integer-valued invariants and systems with soliton solutions accompanied by construction of conserved currents with the homotopy classification. In Section 4, micropolar continua are interpreted in relation with Skyrmions using the measures of defects introduced in Section 1. This relation turns out to be the general case of the discussions in Section 2 using the projective geometry.

We use $\mu,\nu,\rho$ for spacetime coordinates, $i,j,k$ for space indices and $a,b,c$ for internal indices differ from the coordinate labels. We assume indices of vectors are naturally raised and indices of derivatives are naturally lowered, and a metric tensor with its signature $(+1,-1,\cdots,-1)$ in $n$ spatial dimensions with one time component for tensors defined in $(n+1)$-dimensional differentiable manifolds.

\subsection{Micropolar theory}
In the theory of classical elasticity, motion of a body consisting a bulk of particles $P$ can be written as a function of a position vector $\B{x}=\B{x}(\B{X},t)$ in a spatial configuration dependent on the original position $\B{X}$ written in a reference configuration and time $t$, in the usual rectangular Cartesian coordinate system. A displacement vector $\B{u}$ describes an evolution of a point particle at $P$ with a vector $\B{X}$, to a point $p$ with $\B{x}$. This can be written in a spatial description of $\B{u}(\B{x},t)=\B{x}-\B{X}(\B{x},t)$. And a derivative of $\B{u}$ gives rise to a definition of a deformation gradient tensor by
\begin{equation}\label{1.1}
F_{kL}=\frac{\partial x_k}{\partial X_L}\;,\qquad F_{kL}=R_k^{\phantom{k}l}U_{lL}
\end{equation}
where the lower case indices indicate the quantities in the spatial frame and the upper cases are for the quantities in the reference frame. The second expression for $F_{kL}$ in (\ref{1.1}) is written in the form of the classical polar decomposition with a rotation $R$ and a symmetric positive-definite stretch $U$ in three dimensions.

A microcontinuum, pioneered by Cosserat brothers \cite{EC}, is a continuous collection of deformable and stable (indestructible) materials points, i.e., with nonzero determinants of $F_{kL}$. The characteristic aspect of the theory with a microstructure is that we assume the microelement to exhibit an inner structure attached to vectors called \textsl{directors}, which span the internal three-dimensional space. The most general elasticity theories with microstructures contain nine additional degrees of freedom originated from the internal deformations such as microrotations, microcompressions and  microshears. Comprehensive accounts of microcontinuum theories and its developments can be found in \cite{AE1964-1, AE1964-2, AE1}.  

\begin{figure}[!htb]
\[
\parbox{4in}{\includegraphics[scale=0.5]{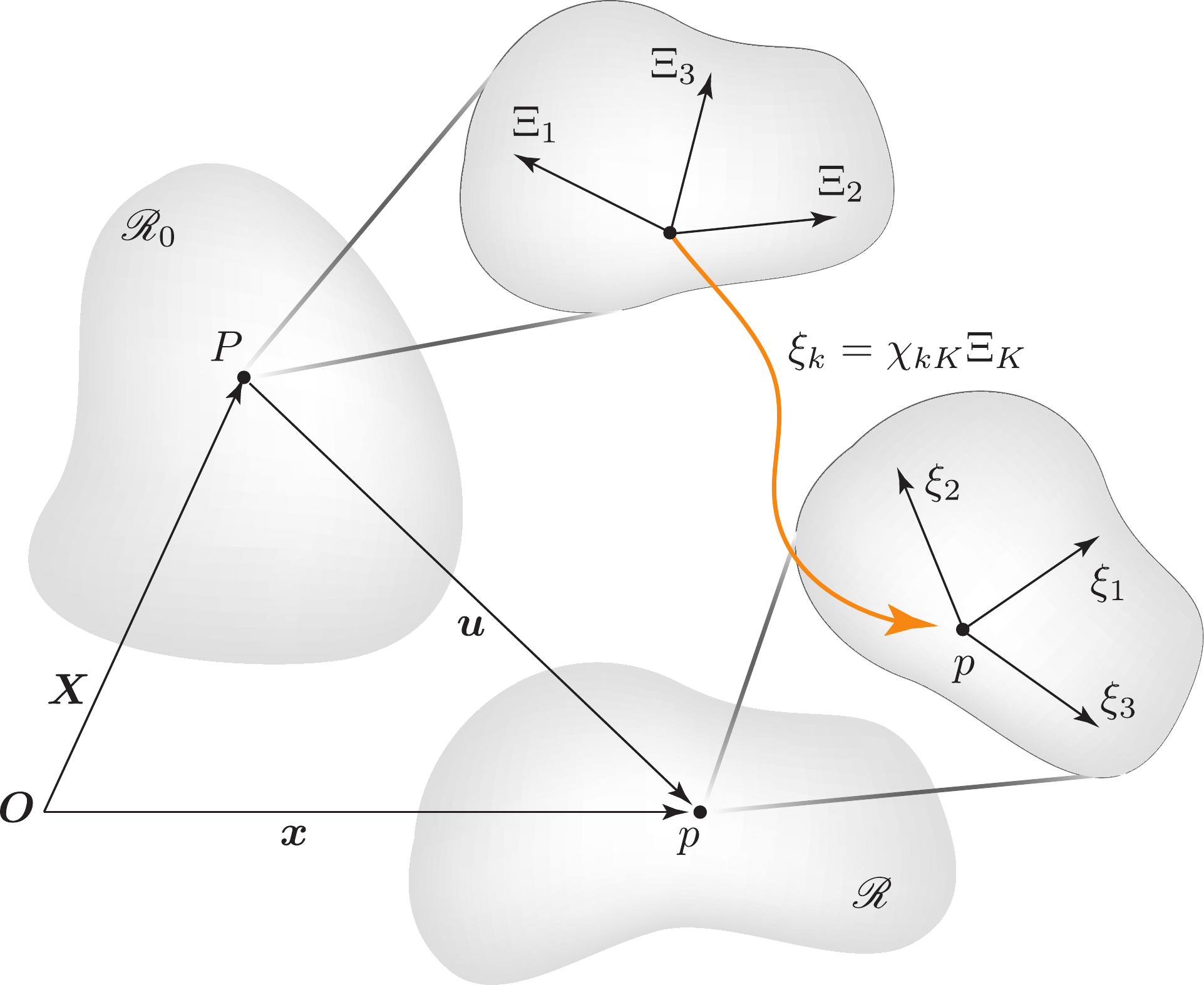}}
\]
\caption{The transformation of the inner structure of the microelement is illustrated with centroids positioned at $P$ and $p$, in the reference configuration and the spatial configuration respectively. This shows how the directors $\B{\Xi}_K$ in the original body in the region $\mathscr{R}_0$ undergoes the microdeformation under $\chi_{kK}$ to become $\B{\xi}_k$ while the original body experiences displacement to become the deformed configuration in the region $\mathscr{R}$ under the macroscopic displacement $\B{u}$.}
\label{f101}
\end{figure}

For the inner structure, we assign a new set of directors $\B{\Xi}_K$ in the reference configuration and $\B{\xi}_k$ in the spatial configuration to describe the microdeformations. So, in addition to the classical elasticity, the transformation of directors $\B{\Xi}_K\to\B{\xi}_k$ is governed by a rank-two tensor $\B{\chi}\in GL(3;\mathbb{R})$, defined by
\begin{equation}\label{1.2}
\B{\xi}_k=\chi_{kK}(X_K,t)\B{\Xi}_K\;.
\end{equation}
If we restrict the general microdeformations to be rigid, one deals with a much simplified model with three degrees of freedom of the microrotation, in addition to the classical translational deformation field. The resulting model is often referred to as the Cosserat elasticity or the micropolar theory, and $\chi_{kK}$ becomes the element of $SO(3)$, see Fig.\ref{f101}.

A solution for the dynamical case for the micropolar elasticity is obtained in \cite{CB2019-1} from the most general total energy functional
\begin{equation}\label{1.3}
V_\text{total}=V_{\text{elastic}}(F,\overline{R})+V_{\text{curvature}}(\overline{R})+V_{\text{interaction}}(F,\overline{R})+V_{\text{coupling}}(F,\overline{R})\;.
\end{equation}
Each individual energy functional is written in terms of the macroscopic deformation gradient tensor $F$ and the microrotation $\overline{R}=\B{\chi}$ which can be conveniently represented by a rotational angle $\Theta$ about a normalised axis $\B{n}_3$,
\begin{equation}\label{1.4}
\overline{R}_{ij}=\cos\Theta\;\delta_{ij}+(1-\cos\Theta)n_{3i}n_{3j}-\epsilon_{ijk}n_{3k}\sin\Theta\;,
\end{equation}
where $\epsilon_{ijk}$ is the totally antisymmetric Levi-Civita symbol in three dimensions. The representation (\ref{1.4}) can be easily translated to the well-known Rodrigues' formula. We will use the overline to denote the microdeformational quantities henceforth, whenever we need to distinguish them.

Equations of motion for the system are obtained from the variational principle of the energy functionals (\ref{1.3}) with respect to $F$ and $\overline{R}$ independently after including kinetic terms. A set of simple ansatz to the system is applied in obtaining analytic solutions, such as i) material points can only experience microrotations about a fixed axis, ii) macroscopic displacements occur along the same fixed axis of the microrotation, and iii) macroscopic elastic displacements and microrotations are both governed by longitudinal deformational waves that propagate with the same wave speed.

Under these assumptions, a set of coupled nonlinear partial differential equations is obtained and further reduced to the equation of motion for the microrotation in a form of the so-called double sine-Gordon equation \cite{PB}
\begin{equation}\label{1.5}
\partial_{tt}\Theta-\partial_{\hat{x}\hat{x}}\Theta+m^2\sin\Theta+\frac{b}{2}\sin 2\Theta=0\;,
\end{equation}
where $m$ and $b$ are real constant parameters, determined by a set of elastic moduli in the individual energy functionals of (\ref{1.3}), and $\hat{x}$ is a rescaled $x$-axis. From this, a microrotational solution $\Theta(x,t)$ is obtained that propagates along the $x$-axis with speed $v$
\begin{equation}\label{1.6}
\Theta(x,t)=4\;\text{arctan}\;e^{\pm k(x-vt)\pm\delta}\;,
\end{equation}
for some constant $\delta$ and a parameter $k$ depend on the set of elastic moduli of $V_{\text{total}}$. In the process of solving (\ref{1.5}), boundary conditions are proposed, based on elasticity considerations for the displacement propagation. These are converted to the boundary conditions for the microrotation propagation,
\begin{equation}\label{1.6-1}
\Theta(\pm\infty,t)\to 0\qquad\text{and}\qquad\partial_\mu\Theta(\pm\infty,t)\to 0\;.
\end{equation}
These elastic boundary conditions emphasise that once the deformational disturbances have passed, the configuration will return to the original one and a point of a deformable body does not experience any deformation when it is far away from the point where the deformation currently occurs.

\subsection{Torsion in Riemann-Cartan manifolds}
In continuum physics, curvature and torsion are based on the sources of two distinct defects, called disclination and dislocation respectively \cite{KK1964, RD1971, MK1977, MK1989}. In developing a theory for the generalised local symmetry under the Poincar\'{e} group in the curved spacetime, the needs for the non-Riemann manifold arise naturally which can contain Cartan's torsion \cite{EC1926}. Essentially the notion of torsion has become evident in completing the theory with the spinning particles coupled to the torsion \cite{RU1956, TK1961, SD1976, FH1976, SH1978}.

Inspired from similarities and its applicabilities in describing the defects in Riemann-Cartan manifolds, links between the theories of continuum physics and the Einstein-Cartan theory were investigated in \cite{HK1987, MK1992, FH2007}. Recent developments can be found in \cite{AY2012, AY2012-1, AY2013, AY2013-1}, including compatibility conditions based on the non-simply connected manifolds and geometrical approaches to the defect theory based on the non-metricity. Since the Riemann curvature tensor satisfies various geometrical identities such as Bianchi's identities, it is natural to expect that these identities also play a role in continuum mechanics if one considers the possibility that the Riemann curvature tensor can contain both measures of pure curvature originated from the metric tensor, and torsion which might arise independently of the metric field. 

Now, the metric tensor emerges as a secondary quantity defined in terms of tetrad fields
\begin{equation}\label{1.8}
g_{\mu\nu}=e^a_{\mu}e^b_{\nu}\delta_{ab}\;,
\end{equation}
where $a,b,c$ are tangent space indices. This tells us that the metric tensor $g_{\mu\nu}$ is obtained from the flat Euclidean metric $\delta_{ab}$ by a set of deformations, governed by an element $e^a_\mu(x)\in GL(N;\mathbb{R})$ at each point $x$ in the given manifold. A dual field to $e^a_\mu$ is defined by $E^\mu_a$ satisfying the relations $e^a_\mu E^\nu_a=\delta^\nu_\mu$ and $e^a_\nu E^\nu_b=\delta^a_b$.

A vanishing metricity $\nabla_\lambda g_{\mu\nu}=0$ is imposed to give rise to the definition of the general affine connection $\Gamma^\lambda_{\mu\nu}$ and the spin connection $\omega_{\mu\phantom{a}b}^{\phantom{\mu}a}$, as a consequence of a covariantly vanishing tetrad $\nabla_\mu e^a_\nu=0$,
\begin{equation}\label{1.8-1}
\Gamma^\lambda_{\mu\nu}=E^\lambda_a\omega_{\mu\phantom{a}b}^{\phantom{\mu}a}e^b_\nu+e^\lambda_a\partial_\mu e^a_\nu\;.
\end{equation}
This general affine connection is not assumed to be symmetric in the lower indices. 

Since any deformation can be regarded as a combination of rotation, shear and dilatation, in the language of the (micro)continuum theory, we can apply the polar decomposition to the tetrad fields similar to that of (\ref{1.1}),
\begin{equation}\label{1.9}
e^a_\mu=R^a_{\phantom{a}b}U^b_\mu\;,\qquad E^\mu_a=R_a^{\phantom{a}b}U^\mu_b
\end{equation}
where $R^a_{\phantom{a}b}$ is a rotation and $U^b_\mu$ is a symmetric positive-definite tensor. Under these decompositions, one finds that the metric tensor (\ref{1.8}) is blind to the rotational field $R$, but only dependents on the stretch $U$.

The Riemann curvature tensor is defined in terms of the general affine connection
\begin{equation}\label{1.10}
R^\rho_{\phantom{\rho}\sigma\mu\nu}=\partial_\mu\Gamma^\rho_{\nu\sigma}-\partial_\nu\Gamma^\rho_{\mu\sigma}+
\Gamma^\rho_{\mu\lambda}\Gamma^\lambda_{\nu\sigma}-\Gamma^\rho_{\nu\lambda}\Gamma^\lambda_{\mu\sigma}\;.
\end{equation}
And a torsion tensor is defined by
\begin{equation}\label{1.11}
T^\lambda_{\phantom{\lambda}\mu\nu}=\Gamma^\lambda_{\mu\nu}-\Gamma^\lambda_{\nu\mu}\;.
\end{equation}
We define a contortion tensor $K^\rho_{\phantom{\rho}\nu\sigma}$ by a difference between the general affine connection and a metric compatible connection $\accentset{\circ}{\Gamma}^\rho_{\nu\sigma}$, also known as the Christoffel symbol
\begin{equation}\label{1.11-1}
\Gamma^\rho_{\nu\sigma}=\accentset{\circ}{\Gamma}^\rho_{\nu\sigma}+K^\rho_{\phantom{\rho}\nu\sigma}\;.
\end{equation}
The contortion satisfies the antisymmetric property $K^{\lambda}_{\phantom{\lambda}\mu\nu}=-K_{\nu\mu}^{\phantom{\nu\mu}\lambda}$. From this, a dislocation density tensor $K$ is defined by  \cite{HK1982, CB2011-1, PN2015},
\begin{equation}\label{1.12}
K_{ij}=\epsilon_{jkl}K_i^{\phantom{i}kl}\;.
\end{equation}

The vanishing Riemann curvature (\ref{1.10}) and its related measures imply the theory is in the regime of elasticity. And a set of partial differential equations may lead to an integrability condition, which is sometimes called the compatibility condition. In \cite{CB2020}, a universal expression of the compatibility condition is studied under the setting of the vanishing curvature tensor in three dimensions. This yields two distinct classes of compatibility conditions, one for the vanishing torsion and another for the non-vanishing torsion. The former is well known by Vall\'{e}e's classical result \cite{CV1992}. This result states that the vanishing Riemann curvature tensor in the deformed body yields the compatibility conditions equivalent to the Saint-Venant compatibility conditions \cite{BB1955, KK1963, SG1972, EK1980, HK1989, SG2003}
\begin{equation}\label{1.12-1}
\Curl\Lambda+\text{Cof}\;\Lambda=0
\end{equation}
where $\Lambda$ is a $3\times 3$ matrix defined by the stretches $U$ and its derivatives. We defined $(\Curl U)_{ij}=\epsilon_{jmn}\partial_m U_{in}$ and $(\text{Cof}\;U)_{ij}=\frac{1}{2}\epsilon_{ims}\epsilon_{jnt}U_{mn}U_{st}$.

The case for the vanishing Riemann tensor but nonzero torsion is known by Nye's result \cite{JN1953}, with an additional condition $U^c_\mu=\delta^c_\mu$ in (\ref{1.9}) 
\begin{equation}\label{1.15}
\Curl\Gamma+\text{Cof}\;\Gamma=0\;,
\end{equation}
where Nye's tensor $\Gamma$ is defined in terms of the contortion tensor,
\begin{equation}\label{1.13}
\Gamma_{ij}=-\frac{1}{2}\epsilon_{ikl}K^{k\phantom{j}l}_{\phantom{k}j}\;.
\end{equation}

Two compatibility conditions (\ref{1.12-1}) and (\ref{1.15}) are shown to be derivable from a universal expression written by the Einstein tensor $G_{ia}$ in three dimensions
\begin{equation}\label{1.14}
G=\Curl\Omega+\text{Cof}\;\Omega\;,
\end{equation}
where the quantity $\Omega_{ai}$ is defined by a contraction of the spin connection $\Omega_{ai}=-\frac{1}{2}\epsilon_{abc}\omega_i^{\phantom{i}bc}$. This derivation is based on the fact that the vanishing Riemann curvature tensor implies the vanishing Einstein tensor in three dimensions. 

There are two important consequences we would like to mention here, under the setting $U^b_\mu=\delta^b_\mu$, which is equivalent to the trivial metric tensor (\ref{1.8}), and the nonzero torsion. The Riemann curvature tensor (\ref{1.10}) can be written entirely in terms of contortions using the decomposition (\ref{1.11-1}),
\begin{equation}\label{1.16}
R^\rho_{\phantom{\rho}\sigma\mu\nu}=\partial_\mu K^\rho_{\phantom{\rho}\nu\sigma}-\partial_\nu K^\rho_{\phantom{\rho}\mu\sigma}+K^\rho_{\phantom{\rho}\mu\lambda}K^\lambda_{\phantom{\lambda}\nu\sigma}-K^\rho_{\phantom{\rho}\nu\lambda}K^\lambda_{\phantom{\lambda}\mu\sigma}\;,
\end{equation}
and, due to the modification in the tetrad $e^a_\mu=R^a_{\phantom{a}b}\delta^b_\mu$, the dislocation density tensor (\ref{1.12}) can be written by the product of (micro)rotation $R$ and its derivative
\begin{equation}\label{1.17}
K=R^T\Curl R\;.
\end{equation}

It is well-known that the existence of dislocations, or equivalently torsion tensors, can be verified by following a small initially closed path in crystal lattice structures to see if the path is broken. In a similar manner, the nonzero curvature can be confirmed  from angular deviations in a set of initially parallely aligned vectors. For this reason, in examining the nonzero torsion in the defects, a set of large field configurations can be classified as the equivalent classes by an assignment between a unit sphere $S^1$ and a space where the microrotations are assumed to be non-trivial. Hence, it is the first homotopy group, also called the fundamental group, for line defects in three dimensions treating $SO(3)$ as its order parameter,
\begin{equation}\label{1.18}
\pi_1(SO(3))\cong\mathbb{Z}_2\;.
\end{equation}
This suggests that we can have two distinct classifications for the compatibility conditions under the vanishing Riemann curvature tensor. One of them is for the trivial class, the classical elastic regime with zero torsion. So that all deformations belong to the identical compatible condition (\ref{1.12-1}), namely a class $\{0\}$. In this class, all configurations can be continuously deformed to the trivial one under the general diffeomorphism. Another classification is the microdeformational description for which we can assign a class $\{1\}$ with the nonzero torsion. In this class, configurations cannot continuously deform into those in the class $\{0\}$. The second classification $\{1\}$ will be the configurations satisfying Nye's compatibility condition (\ref{1.15}). 

We consider a systematic homotopy classification of defects using nematic liquid crystals in some simple cases next.

\section{Nematic liquid crystals}
\subsection{Homotopy of order parameters}
In a sequential representation of a fibre $F$, a total space $E$ and its projected base space $M$ of $F\hookrightarrow E\rightarrow M$, we can express real and complex projective spaces using the Hopf fibrations. We write some of important fibrations for $n$-dimensional spheres $S^n$ as follows
\begin{subequations}\label{2.1}
\begin{align}
S^0&\hookrightarrow S^n\rightarrow\mathbb{R}P^n\;,\\
S^1&\hookrightarrow S^{2n+1}\rightarrow\mathbb{C}P^n\;.
\end{align}
\end{subequations}

Particular interests arise when we consider the homotopy group relation on these fibrations. For example, suppose that the given manifold $M$ is simply connected. Then any simple closed loop contained in the given manifold can be continuously deformed into another loop and eventually can be deformed to a point. Then, by definition of the fundamental group, we will have a trivial homotopy $\pi_1(M)\cong\{0\}$. Since all $S^n$, $n\ge 2$ are simply connected, while $\mathbb{R}P^n$ for $n\ge 2$ are not, we have $\pi_1(S^n)\cong\{0\}$ for $n\ge 2$. Moreover, by the \textsl{Lifting Properties of the fundamental group} \cite{AH2002} between the non-simply connected space and its universal covering space, there is an isomorphism
\begin{equation}\label{2.2}
\pi_n(S^n)\cong\pi_n(\mathbb{R}P^n)\cong\mathbb{Z},\qquad n\ge 2\;.
\end{equation}
This will be useful when one considers the homotopy of an order parameter space $M\cong\mathbb{R}P^n$. Specifically, an \textsl{order parameter space} $M$ can be regarded as an image of a function $\Psi(x)$ for $x\in E$,
\begin{equation}\label{2.3}
\Psi:E\longrightarrow M\;.
\end{equation}
As the system undergoes some phase transitions, either by an external factor or spontaneously, the symmetry $G$ in $M$ may be altered to be its subgroup $H$. Consequently, there may be regions where the degrees of the order is not uniquely defined. These regions are characterised by a modified quotient group $G/H$. These regions are called the \textsl{defects} and the names of defect with respective dimension $d$ are i) monopole: a point-like defect in $d=0$ ii) vortex: a string-like defect in $d=1$ iii) domain wall: a sheet-like defect in $d=2$. These defects can be understood in connection with topological invariant quantities and can be found in diverse physical systems with order parameters describing the defects of distinct nature \cite{GT1976-1, NM1979, MK1988, MK1989, JS1992, AU2001, MK2006}. In \cite{GA2020} the connection between the phase transitions that originated from the spontaneous symmetry breaking and those based on the topological nature is studied. These topological invariants are the classification of the defects for a given dimension belonging to one of the equivalence classes given by the homotopy group of the order parameter space $M$. This means that the homotopy classification determines the allowed range of configurations to be deformed continuously within the given equivalence class.

In practice, after we identify the order parameter space $M$ of (\ref{2.3}), in order to determine the homotopy groups, we will proceed according to following steps.
\begin{enumerate}
\item
We identify the dimension $m$ of the manifold $M$ where the medium is defined. This can be different from the dimension of  physical space where the medium is placed.
\item
We take account of the dimensionality $d$ of the physically possible defect.
\item
We identify the $n$-sphere $S^n$ which surrounds the region of defects.
\end{enumerate}
In general \cite{GT1976-1} , the dimension of $S^n$ is restricted by the $d$-dimensional defect in an $m$-dimensional medium and is classified by the homotopy group
\begin{equation}\label{2.4}
\pi_n(M),\qquad n=m-d-1\;.
\end{equation}
This expression can be seen as the defects with dimension $d$ are being measured by a \textsl{probe} of a dimension $n$ of $S^n$ separated by a line. All of them are contained in the manifold of interest with a dimension $m$. We can assign the degrees of the defect from the measure with a ruler $S^n$ a point in the manifold $M$. This will show the continuous deformation from one description of the defect to other in the form of equivalent classes, hence the homotopy group representation.

A particularly intuitive case is that when there is an isomorphism between the order parameter $M$ and $m$-sphere $S^m$. This allows us to investigate the possible class of defects by relatively simple homotopic considerations of counting the number of windings of $S^n_{\text{phy}}$, representing the physical space, over $S^m_{\text{int}}$ representing the space where the order parameter is defined with a possible internal symmetry. The homotopy plays the role of assigning these two manifolds,
\begin{equation}\label{2.5}
S^n_{\text{phy}}\longrightarrow S^m_{\text{int}}\;.
\end{equation}
This will give us an explicit expression $\pi_n(S^m)$ to obtain a clue whether the classification of defects are trivial $\{0\}$, or something else. A less intuitive case is when $M\cong\mathbb{R}P^n$ but the homotopy can be found by using the relations (\ref{2.2}). It is widely known that the real projective space $\mathbb{R}P^2$ can be viewed as a manifold for nematic liquid crystals \cite{LL1986, MK1988, MK1989, MK2006, GA2012}. We would like to see how the notion of directors can be used in topological and geometrical perspectives when we are looking for the classification of defects using the Hopf fibration (\ref{2.1}).

\subsection{Nematic liquid crystals as projective space of $S^2$}
Given the order parameter space $M\cong S^m$, this space can be further reduced to its submanifold if there exists a set of equivalent relations on the sphere. For example, if we can identify two points $\{\B{n},-\B{n}\}$ as antipodals on the sphere $S^2$, with a normalisation condition $\B{n}\cdot\B{n}=1$, then we can write the quotient space using (\ref{2.1}a) for $n=2$
\begin{equation}\label{2.6}
\mathbb{R}P^2\cong S^2/\{\text{antipodal}\}\;.
\end{equation}
The right-hand side of (\ref{2.6}) is topologically equivalent to a hemisphere and we can flatten it to obtain a disk and its boundary. Hence, by following a schematic process \cite{AH2002, JM2000} of removing the redundancy on $S^2$ we can write $\mathbb{R}P^2$ as a union of a disk $D^2$ and its boundary $\partial D^2$. In general, we can regard the real projective space $\mathbb{R}P^n$ as an $n$-dimensional disk $D^n$ with the ideal points on the boundary $\partial D^n\cong S^{n-1}$, so that we can write
\begin{equation}\label{2.7}
\mathbb{R}P^n\cong D^n\cup\partial D^n\cong D^n\cup S^{n-1}\;.
\end{equation}

There is an additional important feature in the projective space. The projective space can be \textsl{non-orientable}, which is equivalent to say that it may contain a M\"{o}bius band. This is because after we identify the antipodals on the disk, we can cut the cylindrical portion of $D^2\cup\partial D^2$ half, and then half twist to match the antipodals to form a M\"{o}bius band. This gives us the most compact topological representation of the real projective space $\mathbb{R}P^2$, by a union of a M\"{o}bius band $M^2$ and a disk $D^2$
\begin{equation}\label{2.8}
\mathbb{R}P^2\cong M^2\cup D^2\;.\
\end{equation}

In the case of nematic liquid crystals, we can take the order parameter as the measure of the degree of alignment among the molecules. We take the director of nematic liquid crystals by a vector $\B{n}_\text{N}$, representing the average direction of the rigid rod-like molecular structure. This suggests the rotational symmetry is broken while the translational symmetry still holds through the symmetry reduction process from the completely random state. Although the molecule might possess an apparently distinguishable head and tail feature, we do not distinguish the directors. i.e., if the molecules are aligned in one direction, then it possesses a discrete symmetry of $\B{n}_\text{N}\to-\B{n}_\text{N}$. We assume that this vector satisfies the normalisation $\B{n}_\text{N}\cdot\B{n}_\text{N}=1$.

We see that the identification of $\B{n}_\text{N}=-\B{n}_\text{N}$ is nothing but the identification of the antipodals on $S^2\subset\mathbb{R}^3$ with outward-directed normalised vectors are attached to it. Further, if we assign a point of the nematic liquid crystals by a map $\Psi:S^2\to M$ of (\ref{2.3}), the order parameter space of the nematic liquid crystals is defined in the projective plane $\mathbb{R}P^2$. For the director fields depend on the position, this emphasises the difference between the physical space $S^2$ bounded by the topological character and the order parameter space $\mathbb{R}P^2$ due to the identification $\B{n}_\text{N}=-\B{n}_\text{N}$. Therefore, using (\ref{2.6}), we can write an expression of the homotopy group
\begin{equation}\label{2.10}
\pi_n(S^2/\{\text{antipodal}\})\cong\pi_n(\mathbb{R}P^2)\;.
\end{equation}
It is worth noting that in \cite{AI2018}, similar observation was made, but from the lattice space of grains, that the discrete symmetry can induce a non-orientable structure.

Now, we follow the prescribed steps in determining the homotopy group. For the line defects, we have $n=1$ in (\ref{2.4}), and this gives the first homotopy group we can work with,
\begin{equation}\label{2.11}
\pi_1(\mathbb{R}P^2)\cong\mathbb{Z}_2=\{0,1\}\;.
\end{equation}
This implies that there exist two distinct classifications of line defects in nematic liquid crystals. The class $\{0\}$ is that one can be continuously deformed into a uniform configuration. The class $\{1\}$ represents the non-trivial defect, a stable vortex, which does not decay into the state of $\{0\}$ class. For the point-like defect, the dimensionality of $S^n$ becomes $n=2$, and the corresponding homotopy group is now
\begin{equation}\label{2.13}
\pi_2(\mathbb{R}P^2)\cong\mathbb{Z}\;.
\end{equation}
This indicates that there are point-like defects in nematic liquid crystals classified by an infinite range of integers.

The idea of linking integers with the homotopy group is a central ingredient to represent the topological invariants in our discussion. The notion of the topological invariance becomes much clearer when we consider boundary conditions in nonlinear $O(n)$ models with field constraints $\B{n}\cdot\B{n}=1$ for $\B{n}\in\mathbb{R}^n$. These boundary conditions are restricted by the finite energy requirement leading to the integer-valued conserved charges.

\section{Topological invariants and conserved currents}
\subsection{Conserved currents, winding numbers and homotopy}
We would like to define conserved currents $J^\mu$ and its associated total charges $Q$ in general $d=n+1$ dimensions. The form of the current is different from the conventionally derived quantities, such as Noether's current, from the continuous symmetry in the Lagrangian of the system leading to the conservation of energy and momentum. The associated topologically invariant total charge $Q$ can be a conserved mass, an electric charge, a magnetic charge or a quantum number depending on the physical models. We investigate the geometrical origin of the intuitive and apprehensible notion of the integer-valued charge $Q$. We are particularly interested in showing the relation
\begin{equation}\label{3.14}
Q=N\;,\qquad N\in\mathbb{Z}
\end{equation}
leading to various consequences and interpretations. The forms of the currents $J^\mu$ might appear \textsl{ad hoc} at first sight but its construction will be justified later within the finite energy requirement.

Let us begin with a normalised $(1+1)$-dimensional configuration $\B{n}_\text{v}$ 
\begin{equation}\label{3.15}
\B{n}_\text{v}=\left(\cos(N\phi(x,t)),\;\sin(N\phi(x,t))\right)\;.
\end{equation}
The current $J^\mu$ is defined by
\begin{equation}\label{3.16}
J^\mu=\frac{1}{2\pi}\epsilon^{\mu\nu}\epsilon^{ab}n_a\partial_\nu n_b\;,
\end{equation}
where $\B{n}_\text{v}=n_a$ for $a,b=1,2$ and $\epsilon^{\mu\nu}$ are totally antisymmetric Levi-Civita symbols in two dimensions. We can see that the current is conserved $\partial_\mu J^\mu=0$ by its construction. The associated total charge $Q$ is defined by an integration of the time component of the current over all space, and can be evaluated by
\begin{equation}\label{3.17}
Q=\int J^0\;dx=\frac{N}{2\pi}\left[\phi(+\infty,t)-\phi(-\infty,t)\right]\;.
\end{equation}
We note that from the conservation equation, the charge $Q$ must be a time-independent quantity, hence it possesses an intrinsic topological property. 

For the finite energy requirement, the total charge must be localised. This means if the amplitude of $\phi(\pm\infty,t)$ increases (or decreases) indefinitely or is not contained in a small oscillation, we cannot expect to have the finite-valued total charge. This will eventually violate the finite energy requirement for the given system, hence appropriate boundary conditions on $\phi(x,t)$ must be imposed in order to obtain physically meaningful solutions.

Now we impose the boundary conditions on $\phi(x,t)$ in such a way that either $\phi(+\infty,t)=\phi(-\infty,t)$ or $\phi(+\infty,t)\ne\phi(-\infty,t)$. For the former, the total charge becomes zero which gives the configurations belong to the class $\{0\}$. For the latter, if we further specify the condition to be $\phi(+\infty,t)-\phi(-\infty,t)=2\pi$, then we obtain $Q=N$, leading to the integer-valued infinite classes. We note that this analysis agrees with the boundary conditions we imposed in the case of the deformational wave propagation in (\ref{1.6}) and the well-known sine-Gordon system. In both cases, since we have the localised soliton solutions, it is natural to expect to have the integer-valued conserved charge while the distinct asymptotic values might impose different interpretation when we consider the elastic deformation.

The identical integer-valued result can be obtained in a static two-dimensional case from the purely geometric interpretation. That is, the integer $N$ is the integration of the total changes in the angular variable $f(\phi)$ for $\phi=\phi(x,y)$ along the simple closed contour $C$ divided by $2\pi$, which is the genuine and intuitive notion of the counting the winding number
\begin{equation}\label{3.19}
N=\frac{1}{2\pi}\oint_Cdf\;.
\end{equation}
The homotopy in this case is a map from $S^1$ to $S^1$, hence identical to the $(1+1)$ dimensional case, the classification is
\begin{equation}\label{3.19-1}
\pi_1(S^1)\cong\mathbb{Z}\;.
\end{equation}
Specifically, we can define the two-dimensional static configuration for the normalised vortex field $\B{n}_\text{v}$ identical to (\ref{3.15}) but now $\phi=\arctan(y/x)$. Then we can evaluate the identical form of the current (\ref{3.16}),  with $i,j$ are spatial indices,
\begin{equation}\label{3.21}
J^i=\frac{1}{2\pi}\epsilon^{ij}\epsilon^{ab}n_a\partial_jn_b=\frac{N}{2\pi}\frac{\hat{\B{n}}_2}{r}
\end{equation}
where $r^2=x^2+y^2$ and $\hat{\B{n}}_2=(x,y)/r$. From this expression, we note that the current is not defined at the origin, in agreement with the definition of the vortex field. And its divergence must be proportional to the two-dimensional Dirac delta function, to write
\begin{equation}\label{3.21-1}
\partial_iJ^i=\frac{N}{2\pi}\; 2\pi\;\delta^2(\B{r})\;.
\end{equation}
Therefore, we will obtain the integer $N$ if we integrate (\ref{3.21-1}),
\begin{equation}\label{3.21-2}
N=\int d^2x\;\partial_i J^i\;.
\end{equation}
Hence, regardless of time-dependent or static configurations, we will obtain the conserved total charge $Q=N$ under the identical form of the current $J^\mu$ for a given dimensionality.

As the natural extension to $(2+1)$ dimensions, retaining both the normalisation condition and the encoded integer $N$, the simplest form of the field configuration can be written with an additional angular variable \cite{EW1976, HW1986} to the field $\B{n}_\text{v}$,
\begin{equation}\label{3.23}
\B{n}_3=\left(\sin\theta\cos N\phi,\sin\theta\sin N\phi,\cos\theta\right)\;,
\end{equation}
where $\theta$ is a polar angle and $\phi$ is an azimuthal angle.  In particular, the static configuration with $N=1$ in (\ref{3.23}) is called the hedgehog field $\B{n}_\text{h}$, introduced by Polyakov \cite{AP1974}.

Next, we would like to consider the mechanism that lies beneath in evaluating the integrations in the arbitrary dimensions to assure the integer-valued $Q$ in accordance with the homotopy classification. In $d=(n+1)$ dimensions, the field configuration can be defined by
\begin{equation}\label{3.41}
\B{n}_d=(\sin\omega_{d}(r)\B{n}_{d-1},\;\cos\omega_{d}(r))\;,\quad d\ge 3\;
\end{equation}
where $\B{r}=(x_1,\cdots, x_d)$, $r=|\B{r}|$ and $\B{n}_2=\B{n}_\text{v}$. We are not restricted to the physical space in defining the field configuration (\ref{3.41}) in the Cartesian coordinate system, but it also can be used in defining the configuration in some abstract internal manifolds that share same topological structure with $\mathbb{R}^d$. 

The field configuration (\ref{3.41}) satisfies $\B{n}_d\cdot\B{n}_d=1$ and each angular function $\omega_d(r)$ imposes boundary conditions. Specifically, all field configurations must approach to a fixed configuration as $r\to\infty$. This fixed configuration is sometimes called a vacuum solution that gives a zero-energy solution. Therefore, the physical solution space can be compactified to the sphere $S^n_\text{phy}$. Consequently the mapping between the field configuration on the sphere $S^n_{\text{int}}$, due to the constraint on the field $\B{n}_d$, and the sphere $S^n_{\text{phy}}$ gives precisely the homotopy classifications
\begin{equation}\label{3.42}
\pi_n(S^n)\cong\mathbb{Z}\;.
\end{equation}

For the general $d$-dimensional case, we can write the conserved current by
\begin{equation}\label{3.39}
J^\mu=\frac{1}{n!\int d\Omega_d}\epsilon^{\mu\mu_1\cdots\mu_n}\epsilon^{aa_1\cdots a_n}n_a\partial_{\mu_1}n_{a_1}\cdots\partial_{\mu_n}n_{a_n}\;,
\end{equation}
where the factor $n!$ comes from an obvious number of permutations. The factor $\int d\Omega_d$ is the area of a unit $n$-sphere $S^{n}$ embedded in the $d$ dimensions. This will give us the topological invariant charge $Q$ by the integration
\begin{equation}\label{3.39-1}
Q=\int d^dx\;J^0\;.
\end{equation}
To see this, for the Euclidean length element $x$, defined by the strictly positive-definite metric tensor, the arbitrary volume element $d^dx$ in (\ref{3.39-1}) can be converted to
\begin{equation}\label{3.38}
d^dx=d\Omega_d\;dx\;x^{d-1}\;.
\end{equation}
Hence the factor $\int d\Omega_d$ in (\ref{3.39}) will be canceled out exactly in the integration. In the static $d$-dimensional Euclidean space, the conserved current is the identical form of (\ref{3.39}), but with the spatial coordinates. The current will take a form of
\begin{equation}\label{3.37-1}
J^i=\frac{N}{\int d\Omega_d}\frac{\hat{\B{r}}}{r^{d-1}}\;,
\end{equation}
which yields a $d$-dimensional Dirac delta function
\begin{equation}\label{3.37-2}
\partial_i J^i=\frac{N}{\int d\Omega_d}\left(\int d\Omega_d\right)\delta^d(\B{r})\;.
\end{equation}
The form of the integration for the charge $Q$ is simply, by using the divergence theorem,
\begin{equation}\label{3.37}
Q=\int d^dx\;\partial_i J^i=\int dS_i J^i\;.
\end{equation}
The factor $dS_i$ on the right-hand side of (\ref{3.37}) is an area of a sphere $S^{d-1}$ in the direction of $J^i$. Hence (\ref{3.37}) gives $Q=N$ identically. In the case of the static configuration, we can obtain the integer $N$ as the winding number from the geometrical consideration on the field configuration $\B{n}_d$ in which the integer $N$ is embedded naturally. In the case of the time-dependent field configuration, the integer $N$ can be obtained from the integration (\ref{3.39-1}) if we impose appropriate boundary conditions for the angular variables, based on the finite energy requirement.

Nonetheless, the field configuration (\ref{3.41}) is not unique for the purpose of obtaining the topological invariants $Q=N$ but it significantly simplifies the task in evaluating the corresponding currents and charges. Other forms of the current can be found in \cite{AB1975, LF1976} based on the lower bound for the finite-energy consideration.

\subsection{Monopoles}
The three-dimensional case in our discussion deserves special attention when one considers monopoles. We would like to reinterpret some of features of the magnetic monopole in connection with the theory of defects we discussed so far. Our approach will highlight advantages in using the field configuration $\B{n}_3$ of (\ref{3.23}) in expressing the associated charge $Q$ in its integration of the current $J^i$, and in visualising the anisotropic field configuration.

In searching for the system that may contain soliton solutions in the hope of the integer-valued charge $Q$, one finds that the existence of soliton solution is severely restricted by the dimension of the system and its constituents \cite{SC1977}. For example, the Yang-Mills theory alone cannot impose the soliton solution in $(3+1)$ dimensions. But if one insists to have a non-trivial topological invariant, one needs to consider a coupled system of gauge vector fields and scalar fields.

The appearance of the gauge field $A^a_\mu$ can be understood from at least two scenarios in the current occasion. First one follows from the requirement of the locally invariant symmetry group with position dependent parameters in the group generator, promoted from the global symmetry. This leads to the minimal prescription of replacing the ordinary differential derivatives by the covariant derivatives. The second case is explicitly shown by Polyakov \cite{AP1974} in the process of removing the possible divergence of the solution $\partial_\mu\phi^a$ in accordance with the finite-energy condition. These cases lead to the identical replacement of
\begin{equation}\label{3.44}
\partial_\mu\phi^a\longrightarrow D_\mu\phi^a=\partial_\mu\phi^a+g\epsilon^{abc}A^b_{\mu}\phi^c
\end{equation}
in the Lagrangian for the coupled system of the Higgs fields $\phi^a$ and gauge vector fields $A^a_\mu$ in $(3+1)$ dimensions, given by
\begin{equation}\label{3.45}
\mathcal{L}=\frac{1}{2}D_\mu\phi^aD^\mu\phi^a-\frac{1}{4}G^a_{\mu\nu}G^{a\mu\nu}-\frac{\lambda}{4}(\phi^a\phi^a-F^2)^2
\end{equation}
where $G^a_{\mu\nu}=\partial_\mu A^a_\nu-\partial_\nu A^a_\mu+g\epsilon^{abc}A^b_\mu A^c_\nu$ and $g,\lambda>0$, $F$ are some real constant parameters. The field configurations are given by
\begin{equation}\label{3.47}
\phi^a=n_{3a}F(r)\qquad\text{and}\qquad A^a_\mu=\epsilon_{\mu ab}n_{3b}W(r)\;.
\end{equation}
where $\B{n}_3=n_{3a}$ are (\ref{3.23}), and the arbitrary radial functions $F(r)$ and $W(r)$ satisfy the boundary conditions as $r\to\infty$,
\begin{equation}\label{3.48}
F(r)\longrightarrow F\qquad\text{and}\qquad W(r)\longrightarrow\frac{1}{gr}\;.
\end{equation}
We note that the Higgs field is in the identical form with axial fields $\B{\Theta}(\B{x},t)=\B{n}_3\Theta$ in the exponential representation of the rotation, $\exp[i\B{\Theta}\cdot\B{L}]\in SO(3)$, where $\B{L}$ are the generators of the rotational group.

The Lagrangian (\ref{3.45}) is invariant under the local $SU(2)$ group and we might expect this contains the electromagnetic field quantities, such as an Abelian $F_{\mu\nu}=\partial_\mu A_\nu-\partial_\nu A^\mu$ under $U(1)$. Since we are dealing with the $(3+1)$ dimensional case with $a,b,c=1,2,3$, we need to consider the following modified form of the currents \cite{JA1975}, differs from (\ref{3.39})
\begin{equation}\label{3.49}
j^\mu=\frac{1}{8\pi}\epsilon^{\mu\nu\rho\sigma}\epsilon^{abc}\partial_\nu n_a\partial_\rho n_b\partial_\sigma n_c
\end{equation}
where we put the fields $n_a$ by  $\phi^a/|F(r)|=\B{n}_3$. Now the associated total charge can be obtained from the integration over the topological density $j^0$,
\begin{equation}\label{3.50}
\begin{split}
Q&=\frac{1}{8\pi}\int d^3x\;\epsilon^{ijk}\epsilon^{abc}\partial_in_a\partial_jn_b\partial_kn_c\\
&=\frac{1}{8\pi}\int d^3x\;\epsilon^{ijk}\epsilon^{abc}\partial_i\Bigl(n_a\partial_jn_b\partial_kn_c\Bigr)\;,
\end{split}
\end{equation}
in which we used $\epsilon^{0\nu\nu\rho}=\epsilon^{ijk}$. The form of the integration in the last line of (\ref{3.50}) is exactly the integration for the derivative of current $J^i$ in the three-dimensional static case of (\ref{3.37-1}). Using the three-dimensional Dirac delta function of (\ref{3.37-2}) it is straightforward to see that $Q=N$. 

Because it manifests a static solution now, it signals that we are allowed to fix the gauge $A^a_0(\B{x})=0$ for all $\B{x}$. This further leads us to obtain the condition for the finite energy requirement in the Hamiltonian. From this observation, we obtain the trivial vacuum solution which gives a zero-energy solution $D_i\phi^a_\text{vac}=0$ and the non-trivial solution which minimises the energy satisfying the boundary conditions (\ref{3.48}).

The Lagrangian (\ref{3.45}) differs from the energy functions of Cosserat elasticity or the nonlinear $O(n)$ models. It includes the gauge field so that $D_\mu\phi^a\to 0$ imposes different meaning from that of $\partial_\mu\phi^a\to 0$ to minimise the energy functional as $r\to\infty$. That is, provided the condition $D_i\phi^a\to 0$ is satisfied, we might have nonzero component of $\partial_\mu\phi^a$ if there exists a cancelling contribution from the gauge field $A^a_\theta$. In other words, $\phi^a$ will tend to the vacuum solution $\phi^a_\text{vac}$ pointing different directions in the internal space. Hence, the physical solution space can be compactified to be $S^2_{\text{phy}}$ and by the normalisation, the internal space is $S^2_{\text{int}}$. Therefore the corresponding homotopy is precisely $\pi_2(S^2)\cong\mathbb{Z}$ of (\ref{3.42}) for $n=2$. This is indeed the homotopy classification for the point-like defect, the monopole, according to (\ref{2.4}).

For the sake of completeness, we show this is the \textsl{magnetic} monopole. Maxwell's equations in the Gaussian unit are given by
\begin{equation}\label{3.55}
\partial_\mu F^{\mu\nu}=4\pi k^\nu
\end{equation}
where $k^\mu$ is the electric current. The dual field is defined by $\widetilde{F}^{\mu\nu}\equiv\frac{1}{2}\epsilon^{\mu\nu\rho\sigma}F_{\rho\sigma}$ which satisfies $\partial_\mu\widetilde{F}^{\mu\nu}=0$. This homogeneous conservation equation of the dual field is the Gauss's law stating that the magnetic flux over the closed surface must vanish.

Following 't Hooft's gauge-invariant definition \cite{GH1974} for the generalised non-Abelian field tensor $\mathcal{F}_{\mu\nu}$, we write
\begin{equation}\label{3.56}
\mathcal{F}_{\mu\nu}=n_aG^a_{\mu\nu}-\frac{1}{g}\epsilon^{abc}n_aD_\mu n_bD_\nu n_c\;.
\end{equation}
Under the configuration (\ref{3.56}), unlike conventional Maxwell's equations, the derivative of the dual tensor $\widetilde{\mathcal{F}}^{\mu\nu}\equiv\frac{1}{2}\epsilon^{\mu\nu\rho\sigma}\mathcal{F}_{\rho\sigma}$ does not vanish, but yields the current (\ref{3.49})
\begin{equation}\label{3.57}
\partial_\nu\widetilde{\mathcal{F}}^{\mu\nu}=\frac{1}{2g}\epsilon^{\mu\nu\rho\sigma}\epsilon^{abc}\partial_\nu n_a\partial_\rho n_b\partial_\sigma n_c=\frac{4\pi}{g}j^\mu\;.
\end{equation}
Comparing with (\ref{3.55}), we conclude that the magnetic current is $j^\mu/g$. Moreover, the total magnetic monopole charge $m$ can be obtained by the following integration,
\begin{equation}\label{3.58}
m=\frac{1}{g}\int d^3x\;j^0=\frac{1}{g}\int d^3x\;\partial_iJ^i=\frac{1}{g}\;N\;.
\end{equation}
Therefore, the non-vanishing current for the monopole can be obtained and the charge is $Q=N$ in the unit of $1/g$. 't Hooft used $N=1$ configuration with the hedgehog field $\B{n}_\text{h}$, and its corresponding solutions under $\lambda\to 0$ are known as the Prasad-Sommerfield solution \cite{MP1975}.

\begin{figure}[!htb]
\begin{tabular}{cccc}
\parbox{1.5in}{\includegraphics[scale=0.3]{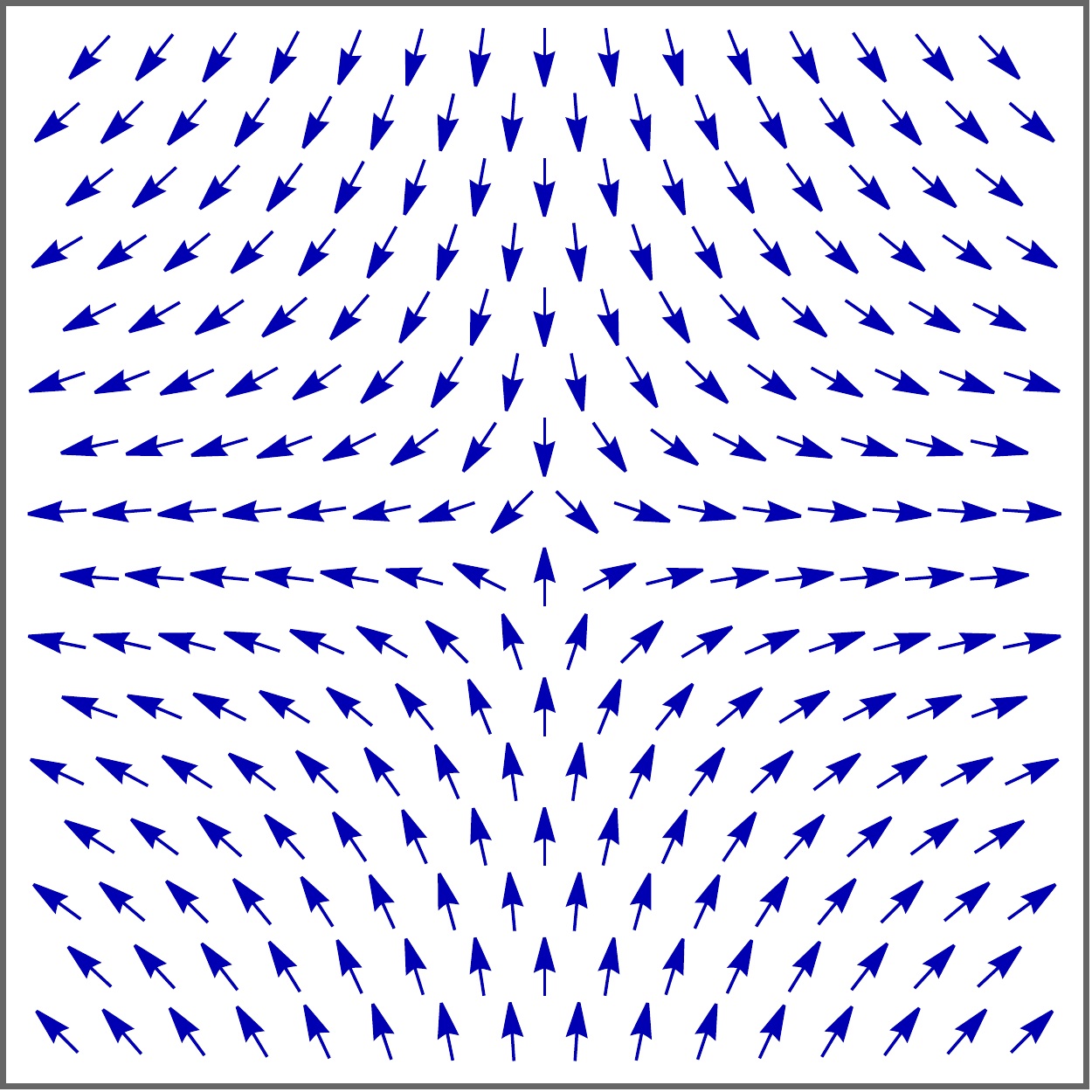}}
&\parbox{1.5in}{\includegraphics[scale=0.3]{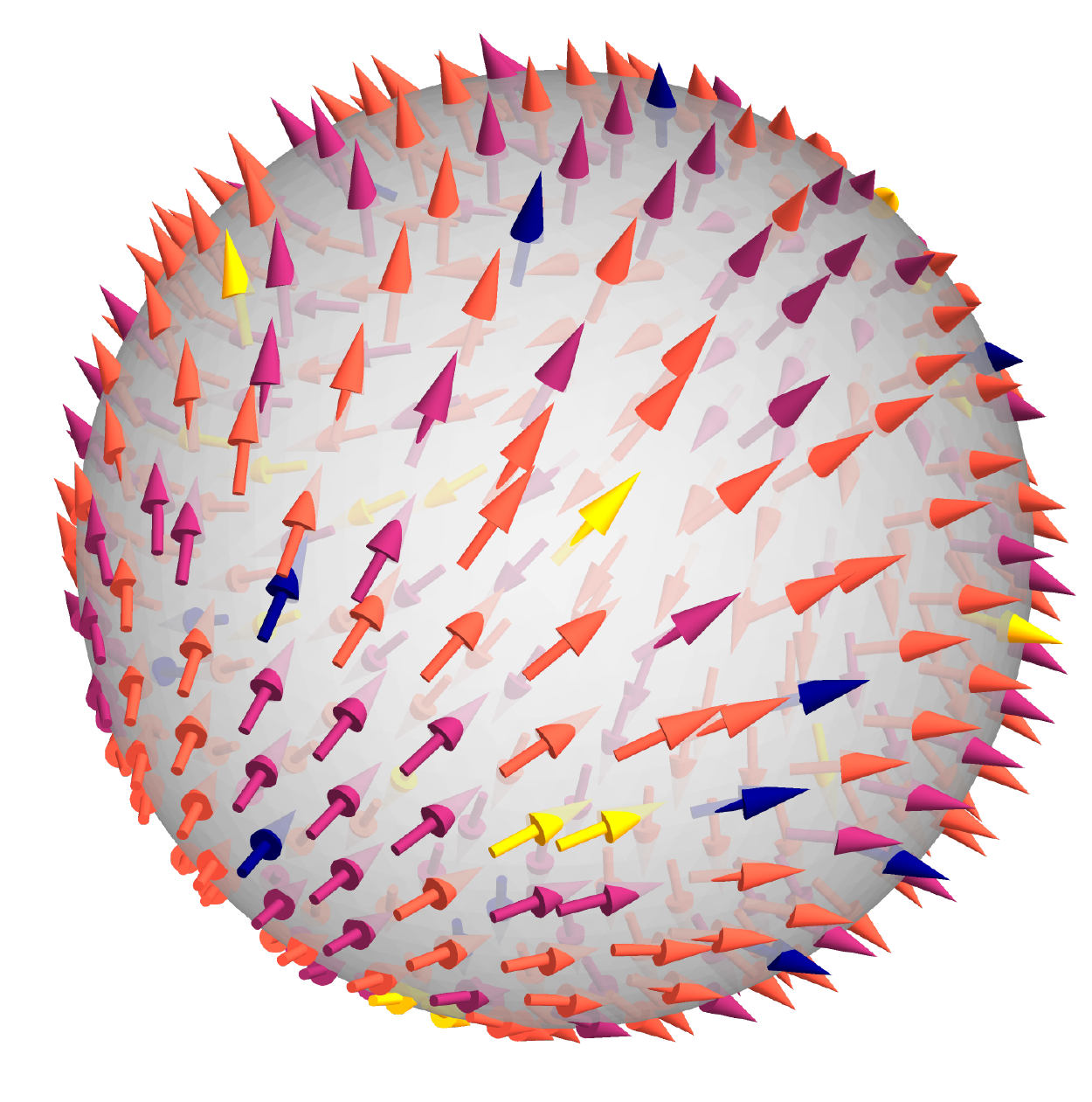}}
&\parbox{1.5in}{\includegraphics[scale=0.3]{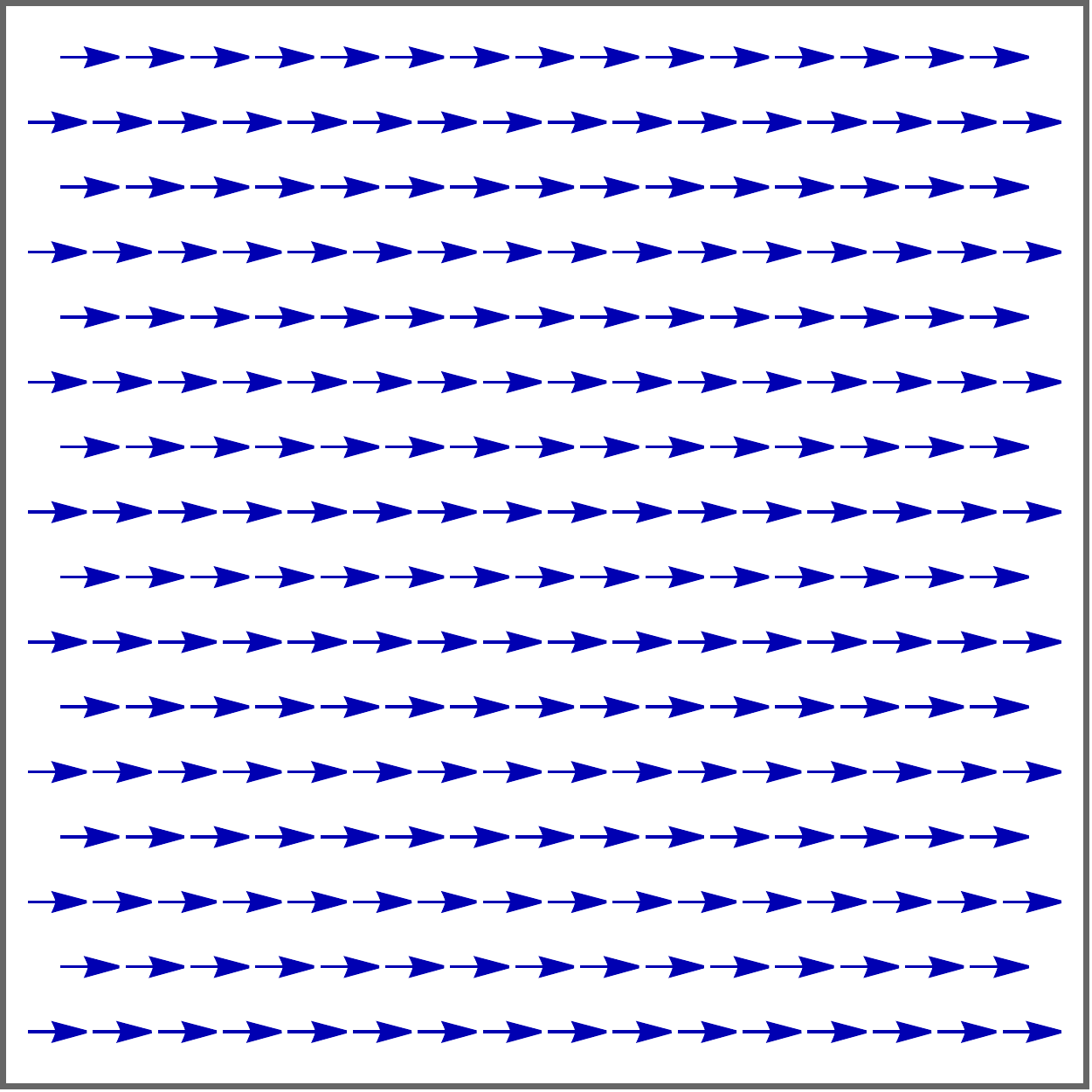}}
&\parbox{1.5in}{\includegraphics[scale=0.3]{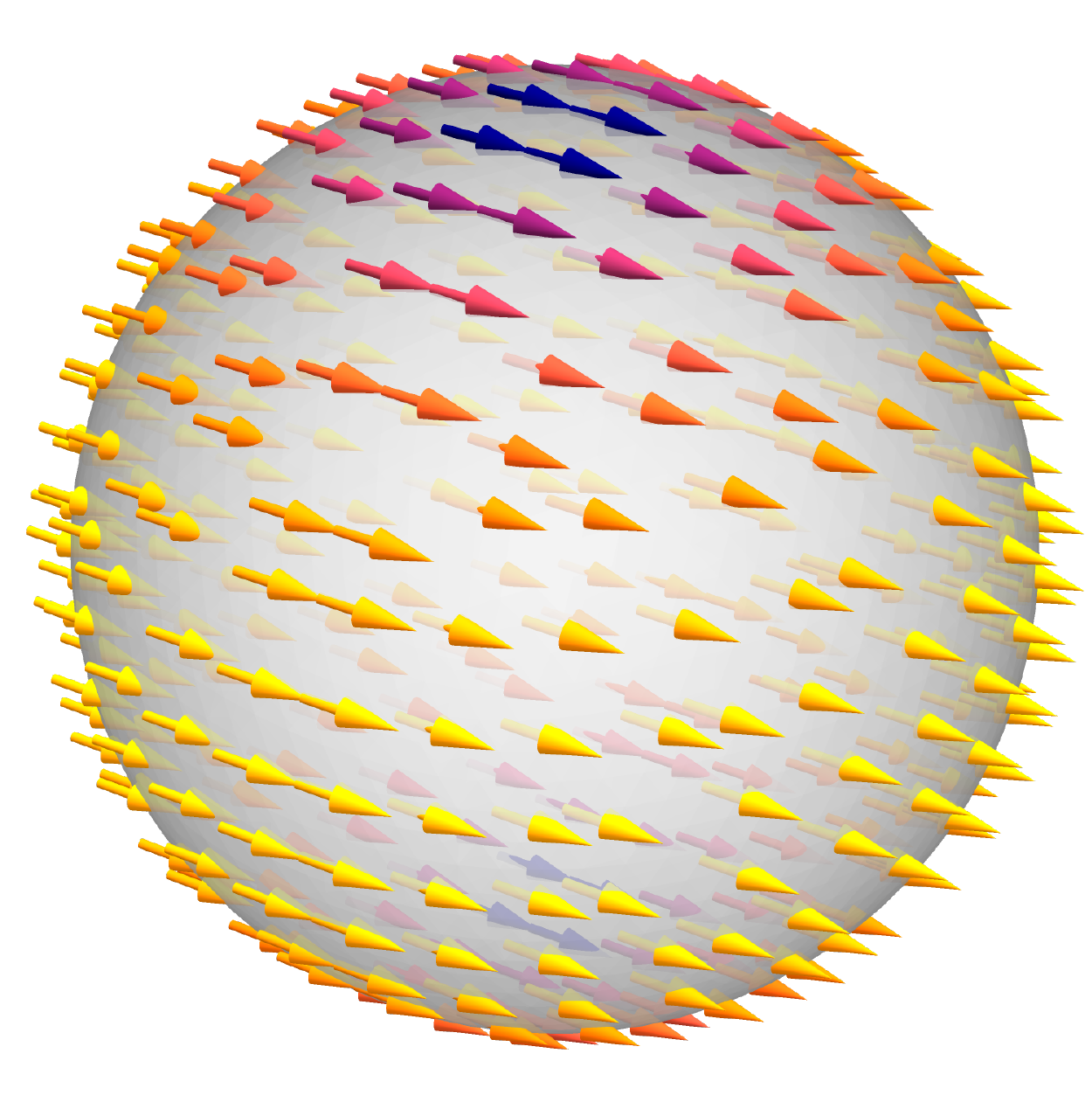}}\\
$\B{n}_\text{v}(N=-1)$&$\B{n}_3(N=-1)$&$\B{n}_\text{v}(N=0)$&$\B{n}_3(N=0)$\\
\parbox{1.5in}{\includegraphics[scale=0.3]{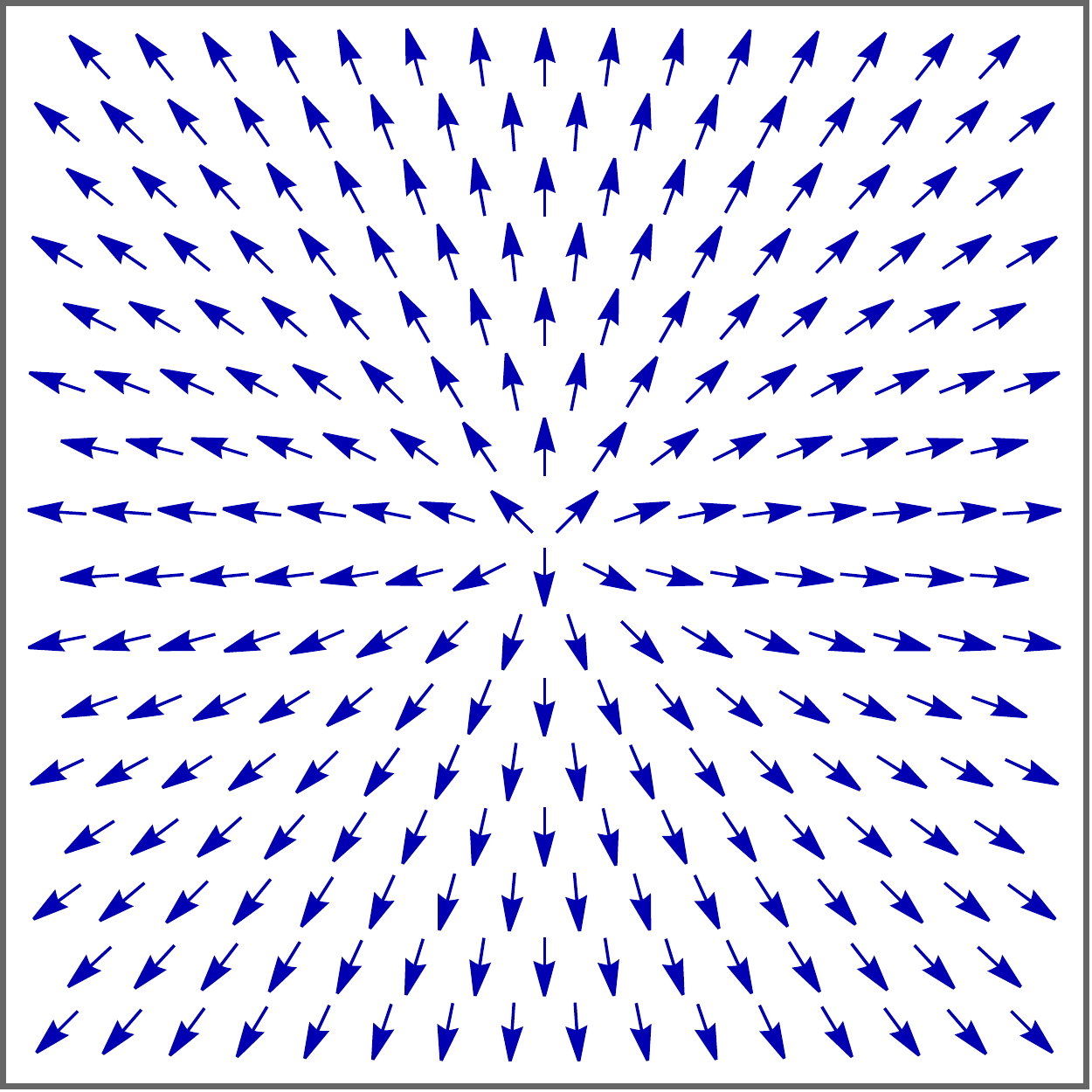}}
&\parbox{1.5in}{\includegraphics[scale=0.3]{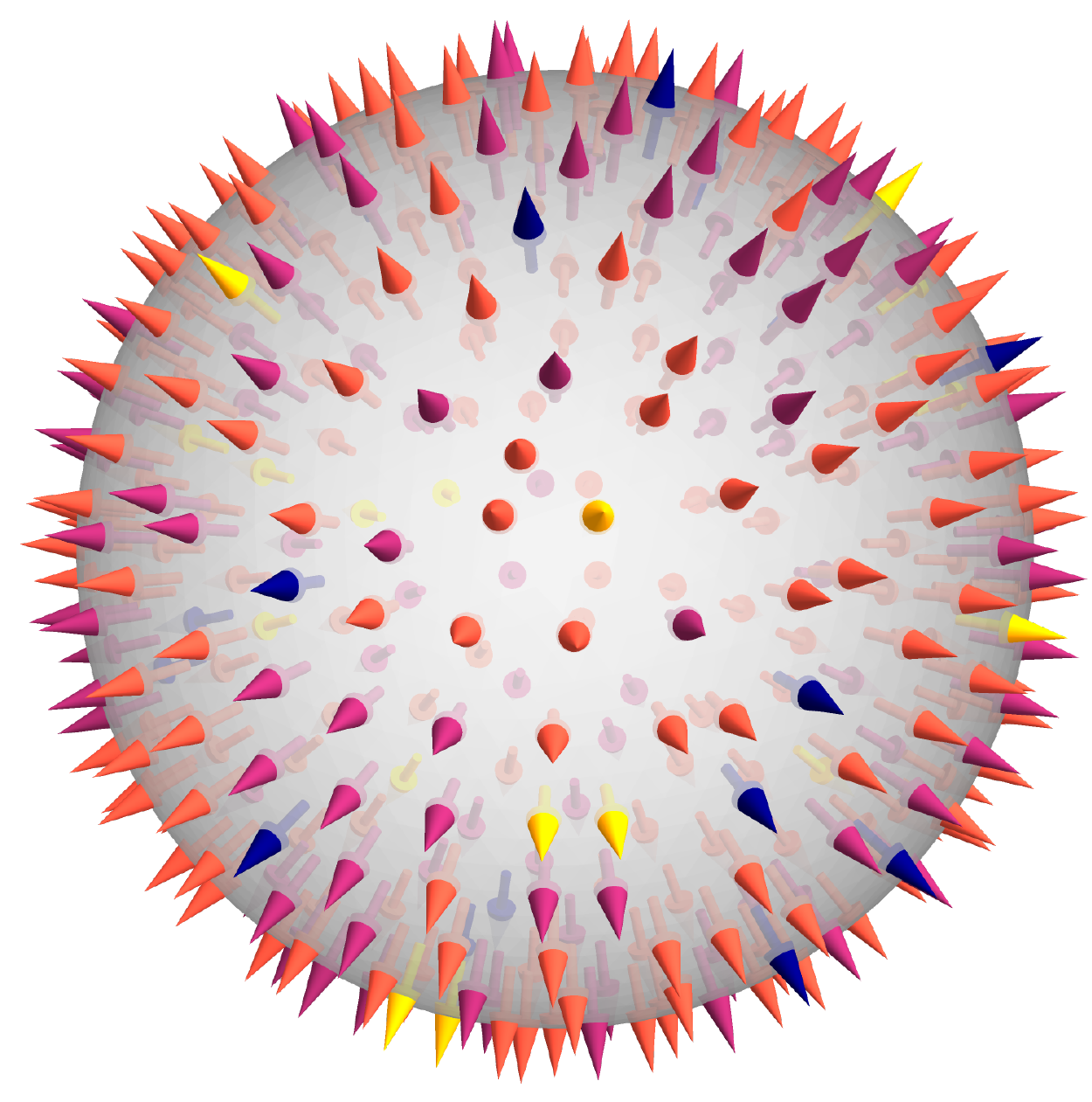}}
&\parbox{1.5in}{\includegraphics[scale=0.3]{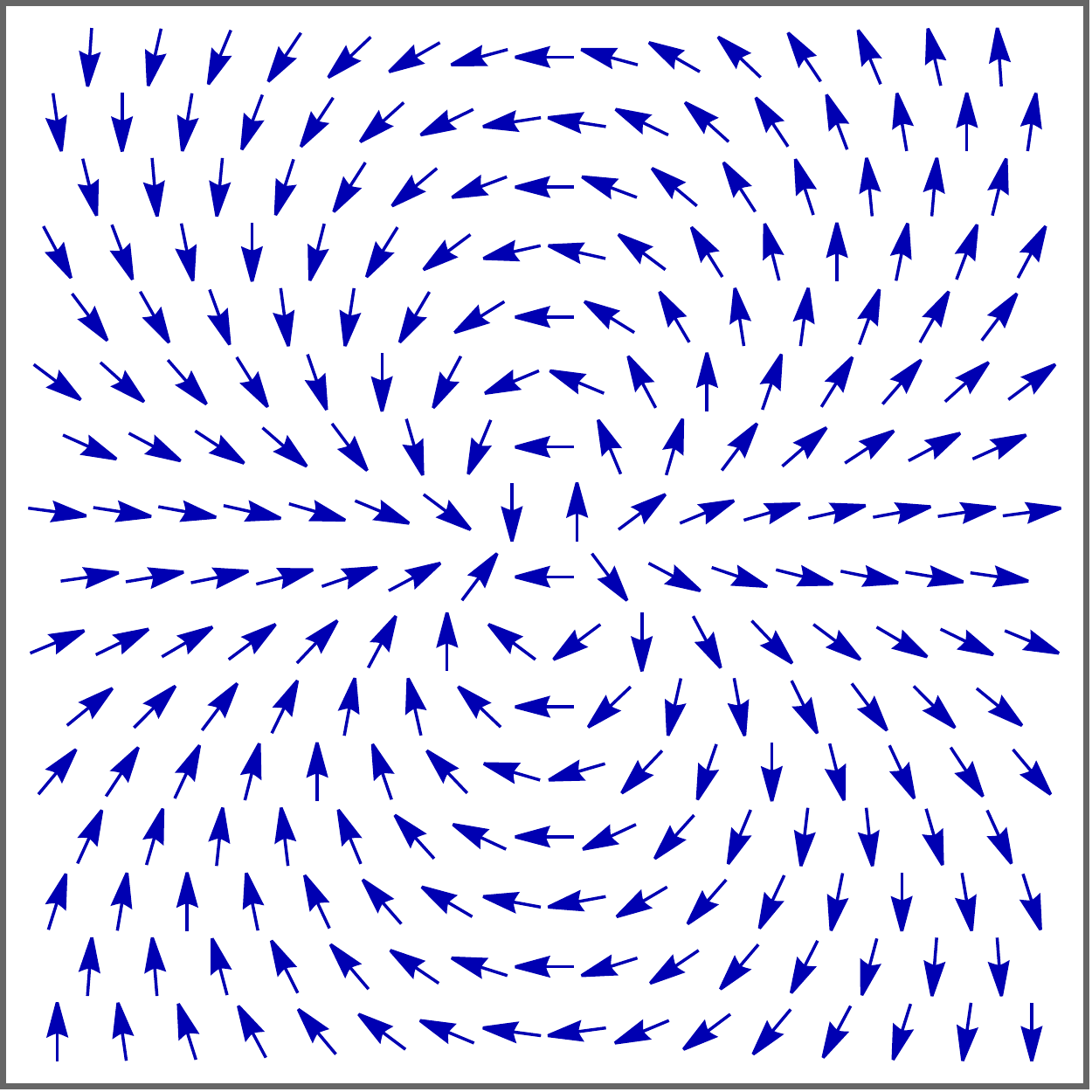}}
&\parbox{1.5in}{\includegraphics[scale=0.3]{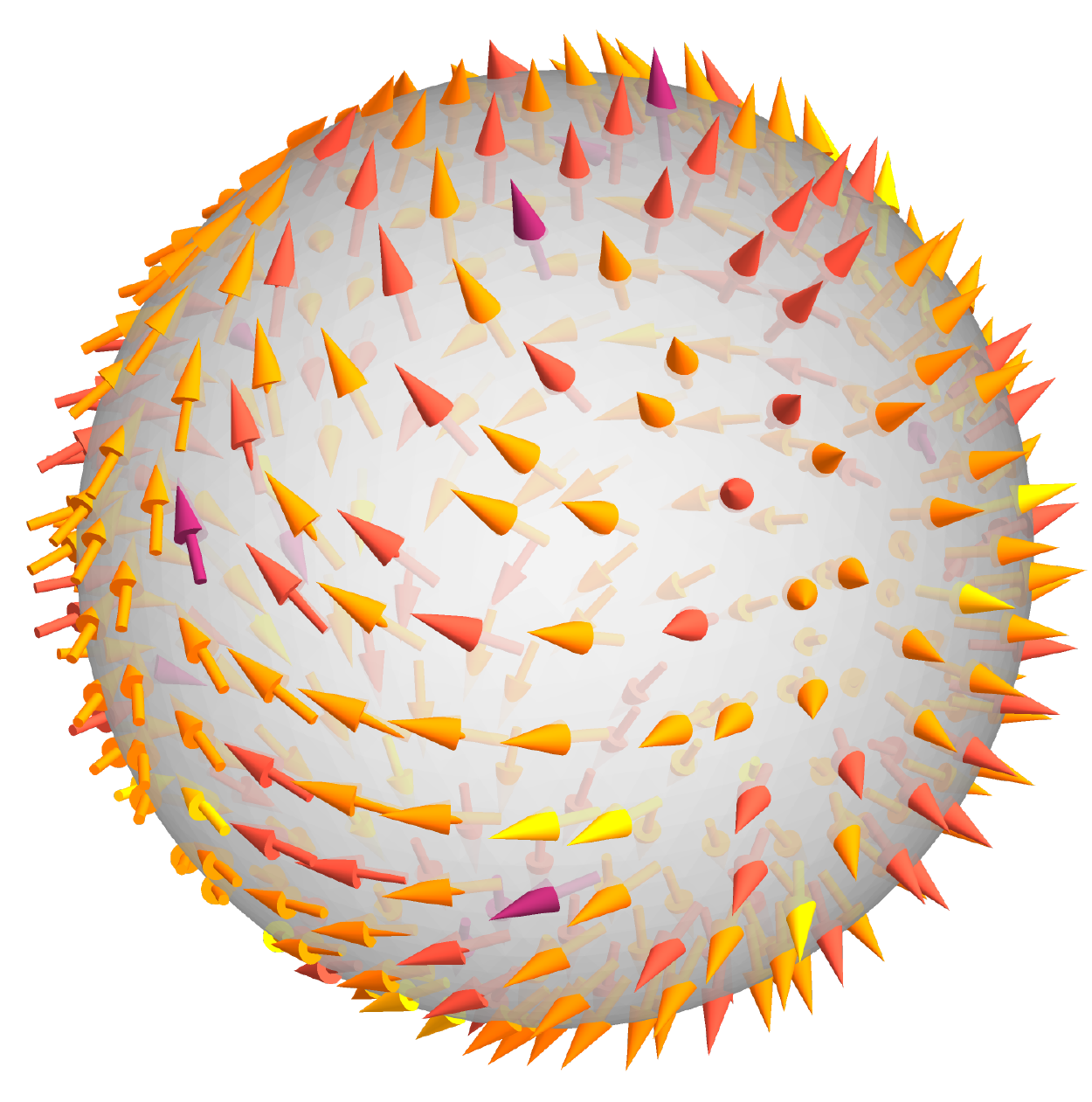}}\\
$\B{n}_\text{v}(N=1)$&$\B{n}_3(N=1)$&$\B{n}_\text{v}(N=2)$&$\B{n}_3(N=2)$
\end{tabular}
\caption{Two-dimensional vortex field configurations with various integers $N$ using (\ref{3.15}) and its corresponding $\B{n}_3$ fields on $S^2$ of (\ref{3.23}) are shown where $\B{n}_3(N=1)$ is essentially the isotropic distribution of the hedgehog configuration $\B{n}_\text{h}$.
}
\label{f301}
\end{figure}

Now, let us consider the vacuum solution corresponding to the $\{0\}$ classification under $r\to\infty$. This must correspond to the field configuration with $Q=N=0$. We use (\ref{3.23}) explicitly with $N=0$ in $\B{n}_3$, rather than fixing the configuration in an arbitrary direction in finding the $\phi^a_\text{vac}$ configuration. This gives,
\begin{equation}\label{3.59}
\phi^a_\text{vac}=F(1,0,0)\;.
\end{equation}
In the region where $\phi^a_\text{vac}$ is defined, the generalised field $\mathcal{F}^{\mu\nu}$ becomes the usual description for the Abelian electromagnetism under $U(1)$, $\mathcal{F}_{\mu\nu}=\partial_\mu A^1_\nu-\partial_\nu A^1_\mu$. Since $Q=N=0$, no monopole can exist in the region where $\phi^a_\text{vac}$ is defined. Moreover, if a field configuration $\phi^a$ belongs to the $\{0\}$ classification, then it must be of the form $\phi^a=\phi^a_\text{vac}+\phi^a_f$ for some small fluctuation $\phi^a_f$ so that $\phi^a$ can continuously deform into $\phi^a_\text{vac}$. This is the transformation that moves the field configuration $\phi^a$ towards the region in which Maxwell's equations (\ref{3.55}) are well defined, see Fig. \ref{f301}.

In an $N=1$ configuration, the field $\phi^a$ is pointing radially outward according to the hedgehog configuration. The monopole, belongs to the $\{1\}$ classification, cannot decay into the vacuum under the finite-energy condition. This again confirms that the magnetic monopoles arise in the static case in which we fixed the gauge $A^a_0=0$ with $F_{0i}=0$ hence no electric fields are defined in the $Q\ne 0$ sector. 

The field $\phi^a$ either can point some fixed direction or can be continuously deformed into the nearby configuration, without violating the finite energy within the given $Q=N$ sector in the range of a small fluctuation. In this respect, this classification is comparable to the sine-Gordon system and the microrotational deformation (\ref{1.6}) in which the different $Q$ sectors correspond to the different values that the field could take asymptotically. In Fig. \ref{f301}, we compare the field configurations for $Q=-1,0,1,2$ cases, arise from the vortex field $\B{n}_\text{v}$ of (\ref{3.15}) embedded in $\B{n}_3$ of (\ref{3.23}). We notice that when $N\ne 1$, the field configuration becomes an anisotropic distribution of $\B{n}_3$. In \cite{EW1976} it is shown that the hedgehog configuration $\B{n}_\text{h}$ is the only isotropic configuration for the magnetic monopole that complies the finite energy requirement.

Although the magnetic monopole we considered here includes the gauge fields to yield the covariant derivatives in the asymptotic behaviours, the construction of the conserved currents is not affected by the presence of gauge fields. This is because the currents are not originated from the Lagrangian but from the intrinsic nature.

\section{Micropolar continua and Skyrme's model}
We would like to focus on the field configuration $\B{n}$ itself constituting the intrinsically conserved current $J^\mu$.  This will include Skyrme's model as the spinor system in which the relation between $SU(2)$ and $SO(3)$ signifies its role in various forms of the representation. It is hinted in \cite{GC1989} that the order parameter for the micropolar continua can be taken as $SO(3)$ or $\mathbb{R}P^3$, and the elements of $R\in SO(3)$ are mentioned briefly in \cite{RS1977} as the antipodals on $S^3$. In this Section, we show that the topological and geometrical generalisation of nematic liquid crystals are micropolar continua. This approach is different from that of \cite{AE1993}.

\subsection{Rotations of $SO(3)$ and $SU(2)$}
An element $U$ of a unitary rotational group $SU(2)$ can be represented by a complex $2\times 2$ matrix, defined by
\begin{equation}\label{4.1}
U=
\begin{pmatrix}
n_4+in_3&in_1+n_2\\
in_1-n_2&n_4-in_3
\end{pmatrix}
=n_4\cdot I+i(\B{n}\cdot\B{\sigma})\;.
\end{equation}
where $\B{\sigma}=(\sigma_1,\sigma_2,\sigma_3)$ are Pauli matrices, $I$ is a $2\times 2$ identity matrix and the real parameters $\B{n}_4=(n_1,n_2,n_3,n_4)$ can be defined according to (\ref{3.41}),
\begin{equation}\label{3.33}
\B{n}_4=\left(\sin\omega(r)\B{n}_3,\;\cos\omega(r)\right)\;.
\end{equation}
Now, the normalisation constraint $\B{n}_4\cdot\B{n}_4=1$ can be justified by the unitary condition of $U^\dagger U=I$, which also states that $\B{n}_4$ is defined on $S^3$. Using the properties of Pauli matrices and $\B{n}_4$ of (\ref{3.33}), the representation (\ref{4.1}) can be translated into
\begin{equation}\label{4.4}
U=\exp\left[i\omega(r)(\B{n}_3\cdot\B{\sigma})\right]=\cos\omega(r)\cdot I+i(\B{n}_3\cdot\B{\sigma})\sin\omega(r)\;.
\end{equation}

An explicit relation between $R\in SO(3)$ and $U\in SU(2)$ can be explained in several ways, but we take the correspondence used by Skyrme \cite{TS1962},
\begin{equation}\label{4.7}
R_{ij}=\frac{1}{2}\trace\left(\sigma_iU^\dagger\sigma_jU\right)
\end{equation}
where $i,j=1,2,3$. A straightforward calculation of (\ref{4.7}) by inserting the matrix elements of $U$ of (\ref{4.1}) gives another representation of $R_{ij}$ in terms of $(n_i, n_4)$,
\begin{equation}\label{4.8}
R_{ij}=2n_in_j-2\epsilon_{ijk}n_kn_4+\delta_{ij}(2n^2_4-1)\;.
\end{equation}
Moreover, by setting $\omega(r)=\Theta/2$ of (\ref{3.33}) and substituting into (\ref{4.8}), we will recover the previous representation (\ref{1.4}).

\subsection{Spinor structure and $2\pi$ rotation}
A system with \textsl{spinors} is characterised by its acquisition of an additional minus sign to its original state after a $2\pi$ rotational transition and returning to its initial state after a full $4\pi$ rotation. This peculiar character of spinors has been observed in many physical systems \cite{RP1960, EN1962, HB1981}, and it is particularly well-known in particle physics \cite{DF1966, HB1967, YA1967, DF1968} that the spin-1/2 particle takes the $4\pi$ rotational transition to return to its original state.

Now, since the correspondence between $SU(2)$ and $SO(3)$ is given by (\ref{4.7}) where each of these are transformations defined on the respective space $S^3$ and $S^2$, there must exist an explicit relation between $S^3$ and $S^2$. Such a relation can be understood by the Hopf fibration. In particular the fibration $S^1\hookrightarrow S^3\rightarrow S^2$ gives a unique identification $\mathbb{C}P^1\cong S^2$ by the  fibration (\ref{2.1}b). For $n=1$ case, we have an isomorphism in terms of complex projective space $\mathbb{C}P^1\cong S^3/S^1\cong S^2$. To see this relation more closely, we define a complex doublet $\B{z}\in\mathbb{C}^2$ which lives on $S^3$. The doublet can be obtained from a state $\binom{1}{0}$, for example, acted by $U\in SU(2)$ of (\ref{4.1})
\begin{equation}\label{4.12}
\B{z}=U\binom{1}{0}=
\begin{pmatrix}
\cos\omega(r)+i\cos\theta\sin\omega(r)\\
ie^{iN\phi}\sin\theta\sin\omega(r)
\end{pmatrix}
\;,
\end{equation}
where we used (\ref{3.33}) for $U$ as a particular example of such a coordinate representation of $\mathbb{C}P^1$. We see that the normalisation condition $\B{n}_4\cdot\B{n}_4=1$ is translated to the condition $\B{z}^\dagger\B{z}=1$.
Hence, we can interpret $\B{z}$ as the wavefunction of spin-1/2 particle field invariant under the $U(1)$ phase transformation $\B{z}\rightarrow e^{i\alpha}\B{z}$, in addition to the originally imposed symmetry $SU(2)$. So, we can recognise the field configuration as the complex projective space $\mathbb{C}P^1$, under the equivalence relation $\B{z}\sim\zeta\B{z}$ for $\zeta\in\mathbb{C}$ and $|\zeta|=1$, as expected. In the case of the real projective space, we take the fibration (\ref{2.1}a) for $n=3$. So that the corresponding quotient space is $\mathbb{R}P^3\cong S^3/\mathbb{Z}_2$, in which we recognise $\mathbb{Z}_2$ the antipodals $\{\B{n},-\B{n}\}$ on $S^3$. Some non-trivial homotopy groups from the fibrations of real and complex projective spaces are
\begin{subequations}\label{4.16}
\begin{align}
&\pi_2(\mathbb{C}P^1)\cong\pi_2(S^3/S^1)\cong\pi_2(S^2)\cong\mathbb{Z}\;,\\
&\pi_1(\mathbb{R}P^3)\cong\pi_1(S^3/S^0)\cong\mathbb{Z}_2\;.
\end{align}
\end{subequations}
We note that (\ref{4.16}b) is the homotopy we considered in (\ref{1.18}) and we see the isomorphism
\begin{equation}\label{4.16-1}
SO(3)\cong\mathbb{R}P^3\cong S^3/\mathbb{Z}_2\;.
\end{equation}

Identifying antipodals with fibrations (\ref{2.1}) can be regarded as the spontaneous symmetry breaking, in some cases of phase transitions that we discussed in Section 2.1. The modified symmetry in the order parameter space $G\to G/H$ will induce a corresponding homotopy group relation $\pi_n(G)\to\pi_n(G/H)$. If $G$ is simply-connected then we can work with much simplified homotopy \cite{SC1988}
\begin{equation}\label{4.16-2}
\pi_2(G/H)\cong\pi_1(H)\;.
\end{equation}
For example, if the original symmetry of the system is $G=SU(2)$ and the reduced symmetry is $H=U(1)$ or $H=SO(3)$ so that the symmetries of the order parameter space of defects are modified to $G/H$, then we have
\begin{subequations}\label{4.16-3}
\begin{align}
&\pi_2(SU(2)/U(1))\cong\pi_1(U(1))\cong\mathbb{Z}\;,\\
&\pi_2(SU(2)/SO(3))\cong\pi_1(SO(3))\cong\mathbb{Z}_2\;.
\end{align}
\end{subequations}
In view of the changes in the symmetry of the system $SU(2)\to SU(2)/U(1)$, the relation (\ref{4.16-3}a) is the homotopy for the region where the magnetic monopole is defined (the nonzero $Q$ sector). And the second relation (\ref{4.16-3}b) is the homotopy (\ref{1.18}). This observation suggests that the compatibility conditions we mentioned in Section 1 might be originated from a larger simply-connected group structure that contains $SO(3)$ as its subgroup, the spinor structure.

Now, the relation between $S^3$ and $S^2$ can be understood in terms of the asymptotic limit $r\to\infty$, especially when we consider the vacuum solutions. For this purpose, the \textsl{Hopf map} $H:S^3\to S^2$ is particularly useful, which gives an explicit expression for the transformations from the complex doublets $\B{z}\in S^3$ into the anisotropic rotational axial fields $\B{n}_3\in S^2$, defined by
\begin{equation}\label{4.17}
H(\B{z})\longrightarrow\B{z}^\dagger\B{\sigma}\B{z}
\end{equation}
where $\B{\sigma}$ are the Pauli matrices. One such coordinate representation of $\mathbb{C}P^1$ is given by an explicit use of (\ref{4.12}). Next, we would like to consider the 3-sphere $S^3\subset\mathbb{R}^4(\cong\mathbb{C}^2)$ and its projection according to the fibration (\ref{2.1}a) in explaining the nature of the boundary conditions for vacuum configuration.
\begin{figure}[!htb]
\[
\includegraphics[scale=0.5]{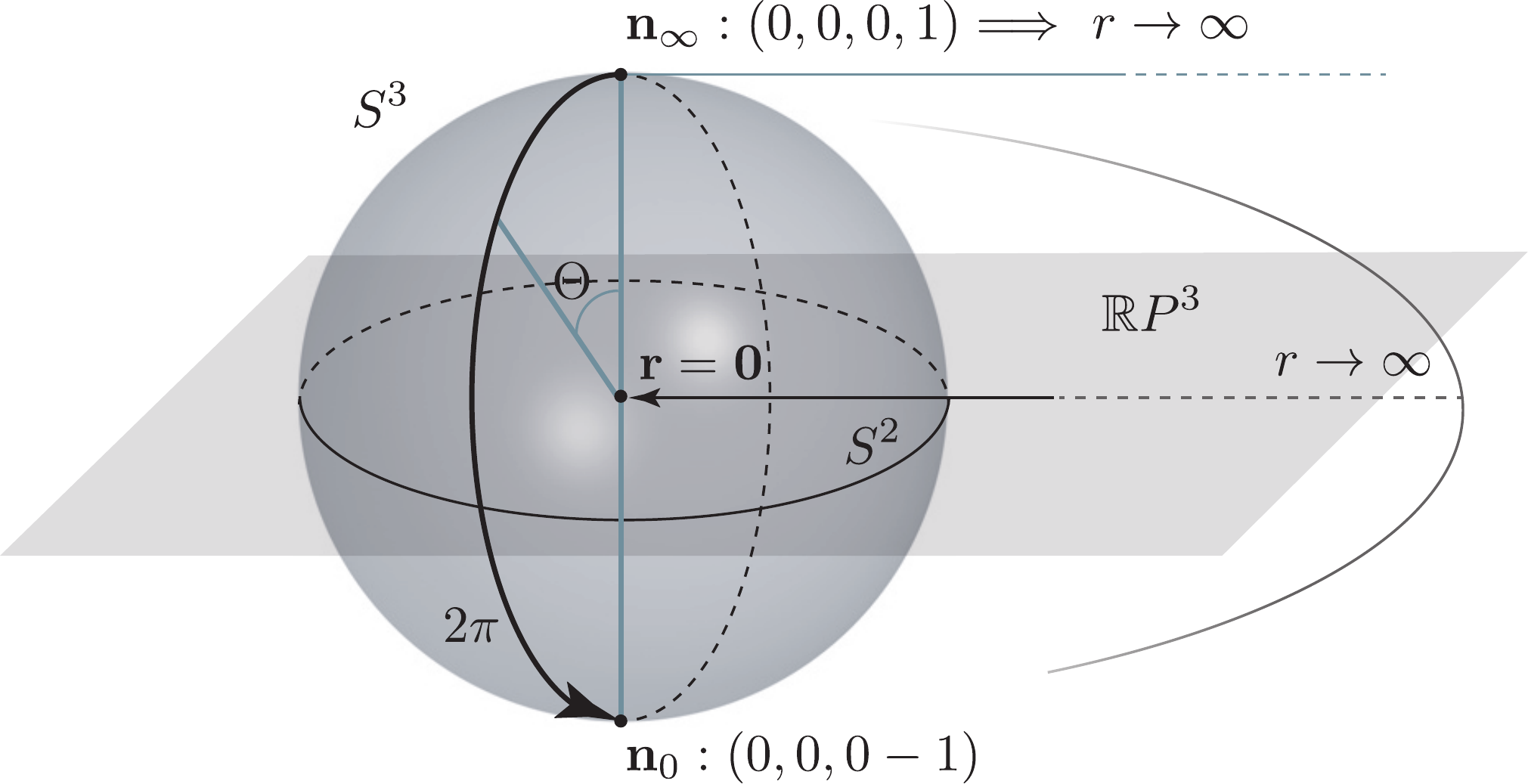}
\]
\caption{The correspondence between $S^3$ and its projection $\mathbb{R}P^3$ is shown with the asymptotic values of $U\in SU(2)$ acting on the point of $S^3$. In particular, the transition of a field configuration starting from  $\B{n}_\infty$ to $\B{n}_0$ is the phase rotation from zero to $2\pi$ on $S^3$. This is a transition that projected on the $\mathbb{R}P^3$ plane by bringing a point from infinity to the origin.
}
\label{f402}
\end{figure}
We consider a map which induces an element $U\in SU(2)$ from a point $\B{r}\in\mathbb{R}^3$ where $\B{r}=(x,y,z)$ is not necessarily normalised vector,
\begin{equation}\label{4.19}
\begin{split}
\B{r}\longrightarrow
\;&\left(\B{\sigma}\cdot\B{r}+i I\right)\left(\B{\sigma}\cdot\B{r}-i I\right)^{-1}\\
=&
\frac{1}{1+r^2}
\begin{pmatrix}
x^2+y^2+(i+z)^2&2(ix+y)\\
2(ix-y)&x^2+y^2+(i-z)^2
\end{pmatrix}
\;.
\end{split}
\end{equation}
It is easy to check that the $2\times 2$ matrix on the right-hand side of (\ref{4.19}) is indeed an element of $U\in SU(2)$. Further, we can restrict the domain $\B{r}$-space as the subspace $\mathbb{R}P^3\subset\mathbb{R}^3$ to construct a correspondence $P:\mathbb{R}P^3\longrightarrow S^3$. And it is easy to see the corresponding representation (\ref{4.19}) satisfies the boundary condition for $U\in SU(2)$
\begin{equation}\label{4.20}
U=
\begin{cases}
 +I\quad&r\to\infty\;,\\
 -I\quad&\B{r}=\B{0}\;.
\end{cases}
\end{equation}

We can defined a field configuration on $S^3$ by using (\ref{3.33}) with $\omega(r)=\Theta/2$
\begin{equation}\label{4.20-1}
\B{n}_4=(\sin\frac{\Theta}{2}\B{n}_3,\;\cos\frac{\Theta}{2})\;.
\end{equation}
Then we can set a coordinate of the north pole on $S^3$ by $\B{n}_\infty=(0,0,0,1)$ for $\Theta=0$ as shown in Fig. \ref{f402}. This corresponds to the point $r\to\infty$ on the projected space $\mathbb{R}P^3$. Now, we apply an element of $U\in SU(2)$ on the state $\B{n}_\infty$ to see the transition by the angular variable $\Theta$, as the phase changes from the north pole to the south pole along a great circle. The south pole denoted by $\B{n}_0$ corresponds to the point $\B{r}=\B{0}$ of $\mathbb{R}P^3$ accordingly. But, by the boundary conditions (\ref{4.20}) and the projective property, we see that the coordinate of the south pole will be $\B{n}_0=(0,0,0,-1)$ with $\Theta=2\pi$.

This means the original state from the north pole acquires a minus sign while the phase transition undergoes the $2\pi$ rotation on $S^3$ along the great circle. From this, the north pole $\B{n}_\infty$ and south pole $\B{n}_0$ on $S^3$ satisfy the asymptotic behaviours on $\mathbb{R}P^3$, while the transition of the element $U\in SU(2)$ changes $+I\to-I$ as the phase changes from 0 to $2\pi$. In particular, when it reaches the middle stage of the transition where $\Theta=\pi$, the state will be on the plane of an equator bounded by $S^2$. Then the field configuration becomes $\B{n}_4=(\B{n}_3,0)$, the field configuration of the normalised magnetic monopole (\ref{3.47}) in four dimensions, see Fig. \ref{f402}.

This establishes a direct link between the state lives on $S^3$ acted by $SU(2)$ and the state lives on $\mathbb{R}P^3$ acted by $SO(3)$ once the antipodals are identified. But what physical system shall we put on $S^3$ and $\mathbb{R}P^3$ acted upon by these rotations respectively to see any physical correspondence? And what is the meaning of \textsl{identifying the antipodals} when one brings an actual physical system to the manifold? We will take such a state on $S^3$ as Skyrmions and we will justify that we can put micropolar continua on the projective space $\mathbb{R}P^3$ next.

\subsection{Skyrmions}
Skyrme \cite{TS1958, TS1959, TS1961-1, TS1961-2} defined the field of the complex doublet $\B{z}\in\mathbb{C}^2$ acted by an element of $U\in SU(2)$ which lies on $S^3$, as we defined in (\ref{4.12}). The construction of this field comes from the fact that two independent $SO(3)$ transformations acting on the intrinsic elementary particle spin space and the isospin space. Skyrme used a pion triplet $(\pi^0,\pi^+,\pi^-)$ where each of this is a meson, the composition of a quark and its antiquark. Then the property of $SU(2)$ as the double cover of $SO(3)$ is used to contain the doublets $\B{z}$. Therefore, two independent full circles in each $SO(3)$ for the spin-isospin coupled field means one $4\pi$ full circle on $S^3$. We note that an isospin space operator has nothing to do with the physical spin space but it acts on the three states of the pion field, and its generators $\B{I}$ share same group structure with that of the generators $\B{L}$ of $SO(3)$.

Skyrme introduced a field $B^a_{\phantom{a}\mu}$ in $(3+1)$ dimensions as a gradient of the pion field, or equivalently a gradient of $U\in SU(2)$ on $S^3$ by
\begin{equation}\label{4.22}
\partial_\mu U=i\sigma^aB^a_{\phantom{a}\mu}U
\end{equation}
where $a,b,c=1,2,3$ are isospin space indices. Then using an identity $\trace(\sigma^a\sigma^b)=2\delta^{ab}$, one can obtain 
\begin{equation}\label{4.23}
B^a_{\phantom{a}\mu}=\frac{1}{2i}\trace\left(U^\dagger\sigma^a\partial_\mu U\right)\;.
\end{equation}
Further, by inserting the representation (\ref{4.8}), directly into the definition (\ref{4.23}), an equivalent expression of $B^a_{\phantom{a}\mu}$ can be obtained by
\begin{equation}\label{4.24}
B^a_{\phantom{\alpha}\mu}=-\frac{1}{4}\epsilon^{abc}R_{bd}\partial_\mu R_{cd}\;.
\end{equation}
Now, let us suppress the index notation for the coordinates $\mu,\nu=1,2,3$ for now. Then we notice that the term on right-hand side $R_{bd}\partial_\mu R_{cd}$ is precisely the form of the contortion tensor $K_{b\mu c}$ of (\ref{1.11-1}) under the condition $U^b_\mu=\delta^b_\mu$, when it is further applied by global rotations $Q\in SO(3)$ according to $Q^TKQ$.

This gives a relation between $B^a_{\phantom{a}\mu}$ field and Nye's tensor (\ref{1.13}) as follows,
\begin{equation}\label{4.25}
B^a_{\phantom{a}i}=\frac{1}{2}\Gamma^a_{\phantom{a}i}\;.
\end{equation}
This relation between two fields gives us a unique identification in what we have discussed in Section 1.3, the compatibility conditions.

We note that since $B^a_{\phantom{a}\mu}$ is in the Maurer-Cartan form by definition, or $i\sigma B=U^\dagger dU$, it must satisfy the Maurer-Cartan equation
\begin{equation}\label{4.26}
dB=-B\wedge B\;.
\end{equation}
After applying Levi-Civita symbols and using the relation (\ref{4.25}), we find that the Maurer-Cartan equation (\ref{4.26}) is precisely our compatibility condition for Nye's tensor (\ref{1.15})
\begin{equation}\label{4.27}
\Curl\Gamma+\text{Cof}\;\Gamma=0\;.
\end{equation}
We recall that this is essentially derived from the vanishing Riemann curvature (\ref{1.16}) but nonzero torsion.

In  \cite{TS1961-1}, Skyrme used an explicit field configuration for (\ref{4.12}) with the components of (\ref{4.20-1}), but the field $\B{n_3}$ is given by the tetrad field $e^a_{\phantom{a}i}$-rotated hedgehog field
\begin{equation}\label{4.31}
\B{n}_3=e^a_{\phantom{a}i}\B{n}_\text{h}\;,
\end{equation}
representing the transformation of the spin-isospin system. The constraint $\B{n}_3\cdot\B{n}_3=1$ immediately indicates that the tetrad field $e^a_{\phantom{a}i}$ must be orthogonal matrices that rotate coordinate and isospin space. On the other hand, the tetrad field $e^a_{\phantom{a}\mu}$ of (\ref{1.9}) rotates coordinate and tangent space, and this will be reduced to the microrotation  $e^a_{\phantom{a}i}=R^a_{\phantom{a}i}$ under the condition $U^b_\mu=\delta^b_\mu$. This might explain the consistent results (\ref{4.27}) and (\ref{4.26}), in two distinct physical systems, which are again equivalent to the vanishing Riemann curvature tensor of the form (\ref{1.16}) expressed by the Maurer-Cartan equation in terms of the contortion $K=R^TdR$,
\begin{equation}\label{4.31-1}
dK=-K\wedge K\;.
\end{equation}

Skyrmions are $(3+1)$-dimensional field configurations for the quantised invariant number defined by the total charge of the integration of conserved current $J^\mu$ defined in (\ref{3.39}) for $d=4$
\begin{equation}\label{4.31-2}
J^\mu=\frac{1}{12\pi^2}\epsilon^{\mu\nu\lambda\rho}\epsilon^{abcd}n_a\partial_\nu n_b\partial_\lambda n_c\partial_\rho n_d
\end{equation}
for $a,b,c,d=1,2,3,4$ and $n_a=\B{n}_4$ of (\ref{3.33}). This topological invariant number is regarded as a particle-like quantity and postulated to be a baryon number. In particular, under the configuration (\ref{4.31}), we obtain the topological invariant charge $Q=N=1$, one proton or neutron. The integer $N=1$ comes entirely from the hedgehog field $\B{n}_\text{h}$. In other words, if we use the general axial configuration such as (\ref{3.33}) or (\ref{4.12}), we will obtain a baryon number $Q=N$ by the following integration of the topological density of the current (\ref{4.31-2}),
\begin{equation}\label{4.33}
N=\int d^3x\;J^0=-\frac{1}{2\pi^2}\int\;d^3x\;\det B\;.
\end{equation}
Using the relation with Nye's tensor (\ref{4.25}), this can be rewritten by
\begin{equation}\label{4.34}
N=-\frac{1}{(4\pi)^2}\int d^3x\;\det\Gamma\;.
\end{equation}
Furthermore, after a rather lengthy calculation using the relation with the contortion (\ref{1.13}), this further becomes
\begin{equation}\label{4.35}
N=\frac{1}{96\pi^2}\int d^3x\;\trace\left(K\wedge K\wedge K\right)\;.
\end{equation}
All three expressions (\ref{4.33}), (\ref{4.34}) and (\ref{4.35}) will give identical topological invariant integer $N$ satisfying the finite energy requirement we considered. The form of the integration (\ref{4.34}) is noted in \cite{HT1981, HT1982} in the context of Cosserat elasticity without referring to the Skyrmions.

When $N=1$, the integration (\ref{4.33}) states that a proton ($Q=1$) cannot decay into the pions ($Q=0$) \cite{TS1961-1, TS1961-2}. We can rephrase this statement by saying that the field configuration belongs to the homotopy class $\{1\}$ cannot continuously deform to be in the class $\{0\}$. And the integration (\ref{4.34}) or (\ref{4.35}) states that the point defects belongs to the non-trivial class $\{1\}$, emphasising the nonzero torsion, differ from that of the class $\{0\}$. 

Therefore, we conclude that the defects in pion fields cannot be measured by means of the metric compatible connection, since it gives only zero curvature in three dimensions leading to the compatibility condition (\ref{4.27}). What remains to describe the non-trivial defects is the nonzero torsion, manifestly links the integrals (\ref{4.33}) and (\ref{4.35}) via the field configuration $\B{n}_3$ for the class $\{1\}$, the isotropic hedgehog field.

It is worth noting that the integration (\ref{4.35}) can be derived from a Chern-Simons type action in terms of the contortion, seen as gauge fields \cite{CB2012-1},
\begin{equation}\label{4.36}
S=\frac{1}{4\pi}\int d^3x\;\trace(K\wedge dK+\frac{2}{3}K\wedge K\wedge K)\;.
\end{equation}
Varying the action $S$ of (\ref{4.36}) with respect to the contortion, one arrives at the equation of motion (\ref{4.31-1}), the vanishing Riemann tensor with nonzero torsion of (\ref{1.16}).

\subsection{Micropolar continua in the projective space}
Let us begin with the fibration (\ref{2.1}a) $S^0\hookrightarrow S^2\rightarrow\mathbb{R}P^2$ for $n=2$. This gives rise to the order parameter space for nematic liquid crystals by identifying antipodals $\B{n}_\text{N}=-\B{n}_\text{N}$ on $S^2$ as we saw in Section 2.2. The natural extension of this consideration would be
\begin{equation}\label{4.38}
S^0\hookrightarrow S^3\rightarrow\mathbb{R}P^3\;.
\end{equation}
We now know the suitable setting for the physical system that lives on $S^3$ is the spinor complex doublet $\B{z}$ of (\ref{4.12}) with the invariants $N$ are embedded in it. There is one remaining problem in understanding the fibration (\ref{4.38}), when it comes to the actual physical model. This is to interpret the geometrical meaning of \textsl{identifying the antipodals} on $S^3$. In case of $S^2$, it comes to the realisation quite intuitively with the aid of the molecular structure of nematic liquid crystals and the relatively simple geometrical property of the director field.

As before, identifying the antipodals will be the statement similar to that of nematic liquid crystals but we put the antipodals to be two identifiable normalised axial field $\B{n}_3$ of (\ref{3.23}), where the antipodals implies $\B{n}_3=-\B{n}_3$ along with the outward-directed rays on $S^3$. Now, we must have an additional degree of freedom to describe the vectors on $S^3$ under the normalisation constraint. The natural candidate for this would be the position-dependent angular variable $\Theta(r)$.

\begin{figure}[!htb]
\[
\includegraphics[scale=0.5]{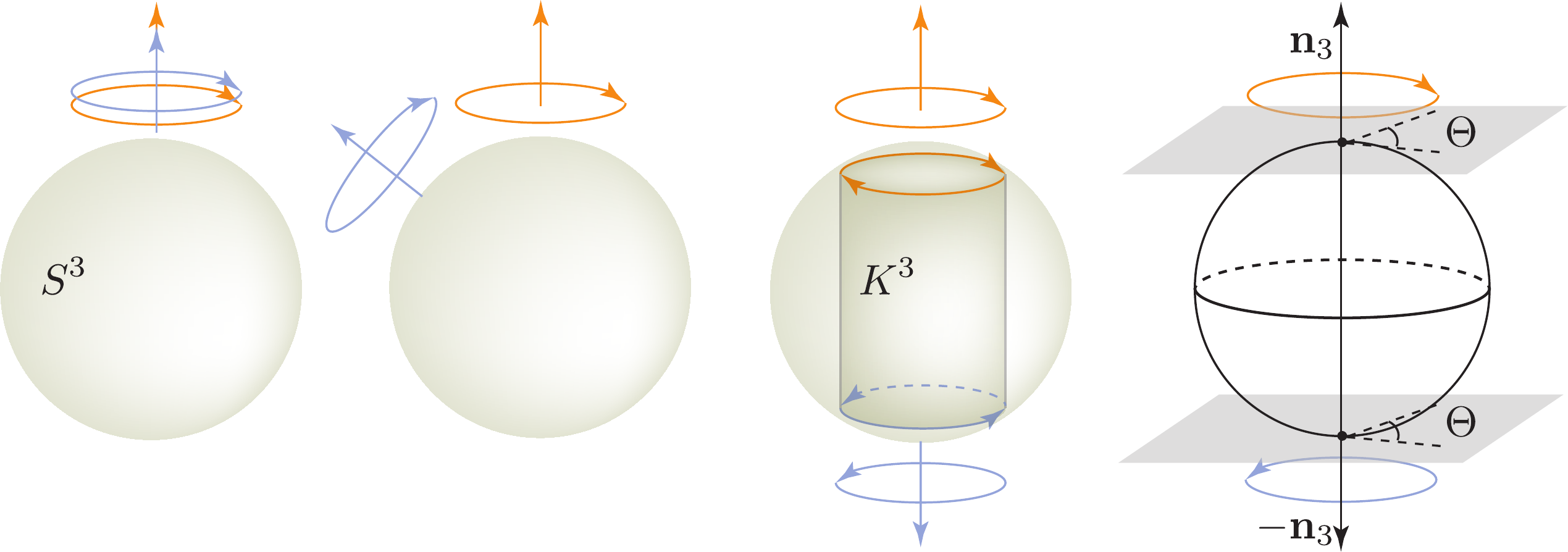}
\]
\caption{Suppose we have started from two states with identical spin orientations on the same axis on $S^3$. As one spinor configuration undergoes a transition along the great circle, separating from the initial configuration which is kept in the initial state, the spin configuration changes gradually. When the phase reaches its $2\pi$ rotation, the spin configuration become complete opposite.
}
\label{f403}
\end{figure}

As shown in Fig. \ref{f403}, suppose a spinor on the north pole $\B{n}_\infty$ of $S^3$ undergoes the angular transition along the great circle $S^2$ with the orientation of the spin by following, for example, a left-hand thumb aligned with the axis of rotation initially. Then  when it reaches the south pole $\B{n}_0$ through $2\pi$ rotation it will acquire an additional minus sign in the assigned vector and we can take these two points on the $S^3$ as the pair of antipodals. We can apply identical analysis on any set of antipodals on the sphere separated by the $2\pi$ rotation along the path $S^2$.

Now, we know that the topological identification of antipodals means that, by the fibration, the quotient space of $S^3/\{\text{antipodals}\}\cong\mathbb{R}P^3$. On the other hand, the geometrical identification of antipodals is equivalent to the statement that the rotation of $2\pi-\Theta$ about $\B{n}_3$ is identifiable to the rotation $\Theta$ about $-\B{n}_3$ as indicated in Fig. \ref{f403}. This is precise the statement of the rotation $R\in SO(3)$. Therefore, the isomorphism is clearly (\ref{4.16-1}). This justifies the identification of antipodals on $S^3$ both geometrical and topological point of views.

In the case of nematic liquid crystals, the angular variable (i.e., the phase transition) has been alway a fixed $\Theta=\pi$ to be restricted on $S^2\subset S^3$. Hence there has been no need for the angular variable consideration but the identification $\B{n}_\text{N}=-\B{n}_\text{N}$ suffices the description for the antipodals on $S^2$. Further, if we consider $R$ as the microrotation, then we can conclude that geometrical identification of antipodals on $S^3$ is the micropolar continuum that lives on $\mathbb{R}P^3$ governed by the microrotational deformation of the angular function $\Theta(r)$ about the anisotropic axial field $\B{n}_3$. Nonetheless, the form of the solution (\ref{1.6}) satisfies $\Theta(\pm\infty)\to 0$, in accordance with Fig. \ref{f402}, but as one approaches the core of the Skyrmion we have $\Theta(0)=\pi$. Hence, our particular solution (\ref{1.6}) in $(1+1)$ dimensions would be suitable for some modified Skyrme's model with its centre may not be at the origin.

We can envision the space of axial fields on the sphere, as space filled with tiny grains rotating independently along with rotational angles about axes determined by position-dependent parameters $\{\Theta(r),\B{n}_3\}$. Once we identify the antipodals on $S^3$, these grains are projected to $\mathbb{R}P^3$ constituting micropolar continua satisfying the boundary conditions we discussed in (\ref{4.20}). Moreover, we can regard the microrotation $\overline{R}_{ij}$ as the order parameter in varying the energy function as we mentioned in Section 1.2,
\begin{equation}\label{4.41}
\frac{\delta V_{\text{total}}}{\delta F}\qquad\text{and}\qquad\frac{\delta V_{\text{total}}}{\delta\overline{R}}\;,
\end{equation}
where $V_\text{total}$ is given by (\ref{1.3}). Since the microrotation $\overline{R}$ can be represented by the angular variable $\Theta(r)$ and the axial vector $\B{n}_3$ we can take the order parameter of the micropolar continuum equipped with the integer $N$ by
\begin{equation}\label{4.42}
Q_{ab}=\left(n_{3a}n_{3b}-\frac{1}{3}\delta_{ab}\right)\Theta(r)\;,
\end{equation}
where $n_{3a}=\B{n}_3$ of (\ref{3.23}) for $a,b=1,2,3$. There is a clear isomorphism between $Q_{ab}$ of (\ref{4.42}) and the microrotational matrix $\overline{R}_{ij}$ represented by (\ref{1.4}). This leads us to the following consequences.

Firstly, the symmetric and traceless matrix form of the order parameter $Q_{ab}$ of (\ref{4.42}) can be viewed as a natural generalisation of that of nematic liquid crystals given in \cite{MK1989, AE1993}. This also agrees with the form of Higgs tensors $\phi^{ab}$, postulated by Polyakov \cite{AP1974}. Then, this implies that if we put $Q_{ab}$ in place of $\phi^a$ in the Lagrangian (\ref{3.45}) in $(1+1)$ dimensions dropping the gauge field, the equation of motion will be of the form
\begin{equation}\label{4.43}
\partial_\mu\partial^\mu Q_{ab}+\frac{\partial V}{\partial Q_{ab}}=0
\end{equation}
where $V$ is the potential. This will be further reduced by applying the normalisation condition on the field configuration $\B{n}_3\cdot\B{n}_3=1$ to yield
\begin{equation}\label{4.44}
\partial_{tt}\Theta-\partial_{\hat{x}\hat{x}}\Theta+\frac{\partial\tilde{V}}{\partial Q_{ij}}=0
\end{equation}
where $\hat{x}$ is the rescaled $x$-axis and $\tilde{V}$ is the modified potential accordingly. Under the isomorphism of $Q_{ab}\cong\overline{R}$, the variation of $V_\text{total}$ with respect to the microrotation in (\ref{4.41}) is just our dynamic equation of motion (\ref{1.5}) yielding the microrotational angle $\Theta$ in $(1+1)$ dimensions. This observation also reinforces the statement that our formulation in deriving the equations of motion is equivalent to the approach from the constitutive equations with the order parameter given in \cite{AE1, CR1990} using the free energy formalism. 

Depending on the form of potential $V$, the equation of motion (\ref{4.44}) can be either the simple $\phi^4$ theory, Klein-Gordon type, sine-Gordon type or more exotic form we encountered in solving the chiral case in \cite{CB2020-1}. In particular, the sine-Gordon system is a special case when we put $b=0$ in the double sine-Gordon system of (\ref{1.5}). This will further reduce to the Klein-Gordon type system when only small angle is allowed. This suggests there might be a direct link between the solution space as the prescriptions of the deformable configuration and the topological invariant quantity $N\in\mathbb{Z}$.

Secondly, identifying the antipodals on $S^3$ corresponds to matching two opposite spin orientations, or to glueing together as shown in Fig. \ref{f403}. But there is another point of view to understand this topological identification, instead of recognising them as the element of $\mathbb{R}P^3$. This is identical to the constructing of a Klein bottle in $S^3$. It is well known that the Klein bottle $K^3$ can be constructed in  $S^3\subset\mathbb{R}^4$ without a self-intersection unlike the usual Klein bottle we are familiar with, and $K^3$ is homeomorphic \cite{JM2000, AH2002} to the connected sum of two projective planes $\mathbb{R}P^2$. This is a generalisation that can be seen from the fact that the gluing antipodals on $S^2$ will lead to the construction of the M\"{o}bius band and the gluing antipodals on $S^3$ will result in the construction of the Klein bottle, see Fig. \ref{f403}.

We know that the projective space $\mathbb{R}P^3$ is a union of $S^2$ and $D^3$ using (\ref{2.7}), and $\mathbb{R}P^2$ is a union of a disk $D^2$ and a M\"{o}bius band $M^2$ of (\ref{2.8}). Therefore, identifying the antipodals on the sphere $S^2$ and $S^3$ implies that we can write the correspondence between the projective space $\mathbb{R}P^3$ and the Klein bottle $K^3$ by
\begin{equation}\label{4.42-1}
\mathbb{R}P^3\longrightarrow\left(M^2\cup D^2\right)\cup\left(M^2\cup D^2\right)\cong
\mathbb{R}P^2\#\mathbb{R}P^2
\cong K^3
\end{equation}
where $\#$ indicates the connected sum along the common disk $D^2$. The relation between a Klein bottle and two (chiral)M\"{o}bius bands can be produced easily by corresponding representations of fundamental polygons.

Although we can only imagine the difficulties in providing the experimental justifications, we summarise the mathematical relations between $S^3$ and $S^2$ concisely. This can be written in the following commuting diagram in terms of the set of morphisms $(H, I, H_p)$ in the category of fibre bundle structures $\mathcal{S}=\{S^3,\mathbb{R}P^3,\mathbb{Z}_2,\pi_1,\psi_1\}$ and $\mathcal{M}=\{S^2,\mathbb{R}P^2,\mathbb{Z}_2,\pi_2,\psi_2\}$ for Skyrmions and magnetic monopoles, bearing in mind micropolar continua and nematic structures are its respective antipodals,
\begin{equation}\label{4.45}
\begin{array}{rccrcr}
\mathcal{S}:&S^3\times\mathbb{Z}_2&\overset{\psi_1}{\xrightarrow{\hspace{1cm}}}&S^3&\overset{\pi_1}{\xrightarrow{\hspace{1cm}}}&\mathbb{R}P^3\\
&(H,I)\xdownarrow{0.7cm}&&H\xdownarrow{0.7cm}&&H_p\xdownarrow{0.7cm}\\
\mathcal{M}:&S^2\times\mathbb{Z}_2&\overset{\psi_2}{\xrightarrow{\hspace{1cm}}}&S^2&\overset{\pi_2}{\xrightarrow{\hspace{1cm}}}&\mathbb{R}P^2
\end{array}
\end{equation}
where $H$ is the Hopf map (\ref{4.17}), $I$ is the identify map, $H_p$ is a restricted Hopf map on the projective space, $\psi_i$ are the group action on the spheres defined by $\mathbb{Z}_2$, and $\pi_i$ are projections from the total space to the base space.

\section{Conclusion and outlook}
We established the firm relation between the nematic liquid crystals on $S^2$ and the projective plane $\mathbb{R}P^2$ by identifying the antipodals in topological and geometrical point of views. As an extension of this idea, we took the physical model on $S^3$ as the Skyrmions based on the recognition that the spin-isospin symmetry for the pion field constitutes the transformation of the spinors acted by $SU(2)$.

This successfully led us to the realisation that the order parameter space of the micropolar continua is the projective space $\mathbb{R}P^3$ when we identify the antipodals on $S^3$, which comes naturally through the correspondence between $SU(2)$ and $SO(3)$. This generalisation is consistent in the topological and geometrical sense accompanied by the symmetric traceless representations of the order parameters for nematic liquid crystals and Higgs tensor respectively.

We considered the criteria for the soliton solution in the framework of the finite energy requirement in connection with the intuitive elastic boundary conditions for the microrotational deformation governed by the soliton solution of the angular variable $\Theta(r)$. This led us the conserved topological invariant as the integer $N$, corresponding to the homotopic classification and the associated conserved charge $Q$ in  arbitrary dimensions.

The vortex field $\B{n}_\text{v}$ with its geometrically characteristic winding number $N$ is used to define the $d$-dimensional field configuration $\B{n}_d$ in the nonlinear $O(n)$ theory. In particular, we showed that the field configuration for the topological invariants is the position-dependent axial field $\B{n}_3$ of the three-dimensional microrotation in which the isotropic hedgehog configuration is the rather special case. This further led us to show that it can be consistently extended to the cases of 't Hooft-Polyakov monopole and Skyrmions. Also it suggests the possible construction of the multipole with anisotropic configurations of $N>1$, and multi-baryon configurations provided individual point-like defects are in the localised configurations of the weakly interacting limit, given the additive nature of the homotopy $\pi_n(S^n)\cong\mathbb{Z}$.

We investigated the 't Hooft-Polyakov magnetic monopole within the scope of the defect classifications by the homotopy. Skyrmions were shown as the description of the defects in the pion field of the nonzero torsion case but with zero curvature, through the compatibility conditions based on the vanishing Einstein tensor in three dimensions. We showed that the defects of the spin-isospin system can be interpreted in the framework of the defects theory in the Riemann-Cartan manifold.

\section*{Acknowledgements}
I am grateful to Christian B\"{o}hmer for helpful comments. This work was supported by EPSRC Doctoral Training Programme (EP/N509577/1).

\bibliographystyle{unsrt}
\bibliography{references}
\end{document}